\documentclass[a4paper,11pt]{article}

\usepackage{amscd,amsmath,amssymb,amsfonts,xspace,mathrsfs,amsthm}
\usepackage{pst-node}
\usepackage{tikz-cd} 
\usepackage{verbatim}
\usepackage{filecontents}
\usepackage{booktabs}
\usepackage{jheppub} % for details on the use of the package, please
\usepackage[T1]{fontenc} % if needed
\usepackage{slashed}
\usepackage{empheq}

\newcommand{\be}{\begin{equation}} 
\newcommand{\ee}{\end{equation}}

\newcommand{\mZ}{\mathbb{Z}}

\newcommand{\mC}{\mathbb{C}}

\newcommand{\mF}{\mathcal{F}}
\newcommand{\CF}{\mathcal{F}}
\newcommand{\mR}{\mathbb{R}}

\newcommand{\mK}{\mathcal{K}}

\newcommand{\mH}{\mathbb{H}}

\newcommand{\mA}{\mathcal{A}}
\newcommand{\mG}{\mathcal{G}}

\newcommand{\mM}{\mathcal{M}}

\newcommand{\rSL}{\textrm{SL}}

\newcommand{\rSO}{\textrm{SO}}

\newcommand{\ep}{\varepsilon}
\newcommand{\mft}{\mathfrak{t}}

\newcommand{\p}{\partial}

\newcommand{\n}{\nonumber}

\title{\boldmath The $i\varepsilon$-Prescription for String Amplitudes and Regularized Modular Integrals 
}

\author{Jan Manschot,}
\author{Zhi-Zhen Wang}

\affiliation{School of Mathematics, Trinity College Dublin, Dublin 2, Ireland \vspace{.2cm}}
\affiliation{Hamilton Mathematics Institute, Trinity College Dublin,\\Dublin 2, Ireland}

\abstract{We study integrals appearing in one-loop amplitudes in string theory, and in particular their analytic continuation based on a string theoretic analog of the $i\varepsilon$-prescription of quantum field theory. For various zero- and two-point one-loop amplitudes of both open and closed strings, we prove that this analytic continuation is equivalent to a regularization using generalized exponential integrals. Our approach provides exact expressions in terms of the degeneracies at each mass level. For one-loop amplitudes with boundaries, our result takes the form of a linear combination of three partition functions at different temperatures depending on a variable $T_0$, yet their sum is independent of this variable. The imaginary part of the amplitudes can be read off in closed form, while the real part is amenable to numerical evaluation. While the expressions are rather different, we demonstrate agreement of the approach with exponential integrals to the contour integral put forward by Eberhardt-Mizera (2023) following the Hardy-Ramanujan-Rademacher Circle Method, and compare these two approaches. We include applications to the Ramond-Ramond sector of the vacuum amplitude and two-point amplitudes of Type I superstring theory.
}

\begin{document} 
	\maketitle
\flushbottom
\newpage
\section{Introduction}
Scattering amplitudes of strings are of central importance to string theory, and may be a way to connect this theory to high energy physics. The $g$-loop amplitude for closed orientable strings with $n$ insertions, $\mA_{g,n}$, takes the form of an integral over the moduli space of string worldsheets with genus $g$ and $n$ punctures, $\mM_{g,n}$ \cite{Green_Schwarz_Witten_2012, Polchinski:1998rq, Polchinski:1998rr, Friedan:1985ge, Moore:1986rh, Verlinde:1987sd}, or related geometries~\cite{Donagi:2013dua, Sen:2014pia, Sen:2015hia}.\footnote{Depending on the string theory and scattering processes, the worldsheet geometry can also involve boundaries or be non-orientable.} The main focus of this paper are one-loop amplitudes, which for closed oriented strings with $n\geq 2$ insertions take the schematic form,\footnote{An $i$ is included in the measure to ensure the measure is real, $id\tau\wedge d\bar\tau=2dx\wedge dy$ for $\tau=x+iy$.

}
\begin{align}
\label{eq:ClosedStringAmp}
 \mA_{1,n}(s_{jk})=\delta^{(d)}\!\left(\sum_{j=1}^{n}p_{j}\right)\int_{\mH/{\rSL(2,\mZ)}}\frac{i\,d\tau\wedge d\bar{\tau}}{({\rm Im}\,\tau)^{2-w}}\,f(s_{jk},\tau,\bar \tau), 
\end{align}
where $s_{jk}=-(p_{j}+p_{k})^{2}$ are Mandelstam variables, $p_j$ are external momenta, $\mathbb{H}/{\rSL(2,\mZ)}$ is the fundamental domain for the torus complex structure $\tau$, and $f$ is a non-holomorphic modular form of weight $w$, which is obtained from the integration of the Koba-Nielsen factor ${\rm KN}_n$ factor~\cite{Polchinski:1998rq} over the $n-1$ non-fixed positions $z_j$,
\be 
f(s_{jk},\tau,\bar \tau)=\int_{(\mathbb{T}^{2})^{n-1}}\left(\prod_{j=1}^{n-1}\,i\,dz_{j}\wedge d\bar z_j\right)\mathrm{KN}_{n}(s_{jk},z_{jk},\bar{z}_{jk},\tau,\bar{\tau}),
\ee 
where $\mathbb{T}$ denotes the torus, and $z_{jk}\equiv z_{j}-z_{k}$. Aside from one-loop amplitudes, integrals over the torus fundamental domain $\mathcal{F}=\mathbb{H}/\rSL(2,\mZ)$ have a range of applications in physics and mathematics~\cite{Petersson1950, Lerche:1988np, Dixon:1990pc, Harvey:1995fq, Borcherds:1996uda, Moore:1997pc, Bringmann2016, Korpas:2019ava}. 
Yet, such integrals are often IR divergent, namely when the $q$-series expansion of $f$,
\be
f(s_{jk},\tau,\bar \tau)=\sum_{m,n} c(s_{jk},m,n)\,q^m\bar q^n,
\ee
involves negative powers, i.e. $c(s_{jk},m,n)\neq 0$ for some $m+n<0$. It is therefore important to develop a suitable analytic continuation or regularization of the integrals, which is among others, important for the question of unitarity in string theory~\cite{Marcus:1988vs, Okada:1989sd, Berera:1992tm, DHoker:1993hvl, DHoker:1993vpp, DHoker:1994gnm, Witten:2013pra, Sen:2016uzq}. An important class of such integrals are those with $n\geq 0$ for all $m+n<0$. These occur for example in threshold corrections of string theory on $K3\times S^1$, where the right-moving sector is supersymmetric \cite{Dixon:1990pc, Harvey:1995fq}, theta lifts \cite{Borcherds:1996uda}, and in correlation functions of observables in the BRST cohomology of topological quantum field theory \cite{Moore:1997pc}. The sum over $U(1)$ fluxes on the Coulomb branch of topological quantum field theory gives rise to a Siegel--Narain theta function similar to those arising for threshold corrections. The divergences for such integrals can be regulated using a by now familiar prescription, namely by integrating first over ${\rm Re}(\tau)$ giving rise to a delta-function $\delta_{m,n}$, removing the terms with negative powers \cite{Dixon:1990pc, Harvey:1995fq, Borcherds:1996uda}. This paper considers more severe divergences where both $m,n<0$. Besides string amplitudes, this type of divergences are also known to occur for certain BRST exact observables in topological quantum field theory \cite{Korpas:2019ava}.

Taking inspiration from the Feynman $i\varepsilon$-prescription in quantum field theory, a string-theoretic prescription for the regularization of such divergent integrals has been developed~\cite{Berera:1992tm, DHoker:1993vpp, DHoker:1994gnm, Mandelstam:2008fa, Witten:2013pra, Eberhardt:2023xck}. This prescription avoids divergences of the integrals by analytically continuation of the integration parameters to the complexification of the moduli space of string worldsheets. One way to view this prescription is that the worldsheet is generally considered in Euclidean signature, except when the string worldsheet develops a long tube in a region of the moduli space. Then for a large value $T_0$ of the proper time parametrizing the tube, the tube worldsheet is Wick-rotated to Lorentzian signature.
For recent applications of the $i\varepsilon$-prescription for string amplitudes, see~\cite{Banerjee:2024ibt,Huang:2024ihm,Yoda:2024qiw}.

In parallel, an alternative regularization for divergent integrals of modular forms over $\CF$ was developed, which was motivated by questions in analytic number theory~\cite{Bringmann2016} and topological quantum field theory~\cite{Korpas:2019ava}. It is a natural question to compare this regularization and the $i\varepsilon$-prescription. 
Indeed, it was observed numerically for the one-loop contribution to the vacuum energy of the bosonic string, that the amplitude $\mathcal{A}^{i\varepsilon}_{\rm closed}$ evaluated using the  $i\varepsilon$-prescription~\cite{Eberhardt:2023xck}, and the amplitude $\mathcal{A}^{\rm r}_{\rm closed}$ evaluated using modular regularized integrals~\cite{Marcus:1988vs, Korpas:2019ava} are identical up to at least seven digits. This paper further explores the connection, and demonstrates using a contour deformation that both prescriptions indeed give identical values, 
\be
\label{eq:ArAeps}
\mathcal{A}^{\rm r}_{1,0}=\mathcal{A}^{i\varepsilon}_{1,0}.
\ee 
The contour deformation depends on the sign of the energy level, but is otherwise straightforward. The contour deformation does apply to more general amplitudes. To illustrate this we evaluate the two-point function for closed string scattering with $s_{jk}=1$ to high precision in Section \ref{sec:2ptAmp}. 

Besides the non-holomorphic integrals over $\mathbb{H}/\rSL(2,\mathbb{Z})$ for closed strings as in Eq. (\ref{eq:ClosedStringAmp}), we also evaluate contour integrals with (weakly) holomorphic integrands arising for open and closed string amplitudes with boundaries using exponential integrals. Such integrals involve both UV and IR divergences due to exponential growth of states or tachyons. For the bosonic open string vacuum amplitude, the $i\varepsilon$-prescription gives rise to the integral~\cite{Eberhardt:2023xck}
\be
\int_{\Gamma(T_0)} d\tau\,\frac{1}{\eta(\tau)^{24}},
\ee 
with the contour $\Gamma(T_0)$ displayed in Figure \ref{fig:fullcontour}. We evaluate this integral resulting in the expression in Eq. (\ref{eq:EvInt}). We discuss the accuracy of numerical evaluation depending on the upper bound $N$ for the truncated sum over mass levels, and the parameter $T_0$. For a given accuracy, we explain the optimal choice for $T_0$ such that $N$ is minimal.

We compare Eq. (\ref{eq:EvInt}) to the evaluation by Eberhardt and Mizera~\cite{Eberhardt:2022zay, Eberhardt:2023xck}, who applied the Hardy-Ramanujan-Rademacher Circle Method  of analytic number theory~\cite{Apostol,Ramanujan:1988,Rademacher:1938}. This approach leads to the contour $\Gamma_\infty$ over an infinite set of Ford circles resulting in Eq. (\ref{eq:ASopen}), such that the integral evaluates to an infinite sum of Ramanujan's sums. It follows from complex analysis that the evaluation of the contour integral in terms of exponential integrals is equivalent to those of~\cite{Eberhardt:2023xck}, which we also verify numerically through explicit computation. The equivalence of the resulting expressions is not manifest at face value, and each have their useful features for convergence and estimates. Section \ref{sec:conc} concludes with a summary of convergence properties and numerical performance of the two regularization strategies.

We derive the general formula  \eqref{equ:genformula} for integrals of this type in terms of Fourier coefficients of the integrands and exponential integrals. Using this formula, we evaluate a few other amplitudes:
\begin{enumerate}
\item zero-point vacuum amplitude for the Ramond-Ramond sector of Type I superstring theory in Section \ref{sec:VacTypeI},
\item planar two-point amplitude of Type I superstring theory in Section \ref{sec:2point}. This amplitude was also considered as a double degeneration limit of four-point amplitudes in~\cite{Eberhardt:2022zay}. The numerical value of Eq. (\ref{eq:2ptvalue}) agrees with~\cite[Eq. (3.31), D.2a and D.3a]{Eberhardt:2022zay}. 
\item non-planar two-point amplitude of Type I superstring theory in Section \ref{sec:np2pt}.
\end{enumerate}
The latter two are relevant for the mass shift and decay rate of string states. For all above amplitudes, we also provide new expressions as exponential sums, Eq. (\ref{eq:RRCircle}), Eq. (\ref{eq:planarCircle}) and Eq. (\ref{eq:NPCircleM}), derived using the Circle Method.

 We leave it for future work to explore more involved amplitudes such as higher-point functions with other external states and higher loop amplitudes. While it will be more involved to evaluate the integrand near the various cusps, we expect that the methods discussed here do carry over.

The outline of this paper is as follows. Section \ref{sec:ieps} reviews the $i\varepsilon$-prescription for string amplitudes. Section \ref{sec:closed1loop} considers the evaluation of zero- and two-point torus amplitudes using the $i\varepsilon$-prescription and regularized modular integrals, with Section \ref{sec:equivalence} proving the equivalence of the two prescriptions. Section \ref{sec:open1loop} considers the bosonic open string vacuum amplitude, and evaluates this in terms of the Fourier expansion of the integrand. Section \ref{sec:FullIntOpen} performs the direct integration and Section \ref{sec:GenInt} provides the general strategy to reproduce the one-loop open string amplitude. Section \ref{sec:TypeIExp} applies the technique to Type I string amplitudes of zero- and two-point functions, both planar and non-planar, and compared with the Circle Method evaluations. Section \ref{sec:conc} concludes with a discussion, highlights the comparison of the two strategies in Table~\ref{tab:Disc} and lists some future directions. In Appendix \ref{app:ModForms}, we review aspects of modular forms and introduce the Rademacher formula for Fourier coefficients, while Appendix \ref{App:C} discusses the analytic behavior and various properties of generalized exponential integrals.
\section{The $i\ep$-Prescription for String Amplitudes}
\label{sec:ieps}
The Feynman $i\varepsilon$ plays a central role in quantum field theory, and it is natural to ask for an understanding within string theory~\cite{Berera:1992tm, Witten:2013pra}.
In $d$-dimensional QFT, the propagators for Euclidean and Lorentzian signatures, are given using the Schwinger parametrization as
\begin{align}
\label{eq:SchwingerProps}
&{\rm Euclidean:}\qquad    \frac{1}{p^{2}+m^{2}}=\int_{0}^{\infty}d\mathfrak{t}_{E}\,e^{-\mathfrak{t}_{E}(p^{2}+m^{2})},\\
\label{eq:SchwingerPropL} &{\rm Lorentzian:}\qquad    \frac{-i}{p^{2}+m^{2}-i\ep}=\int_{0}^{\infty}d\mathfrak{t}_{L}\,e^{-i\mathfrak{t}_{L}(p^{2}+m^{2}-i\ep)}.
\end{align}
where we choose the Lorentzian signature as $-++\dots +$. The Euclidean propagator converges for $p^{2}+m^{2}>0$. The $i\varepsilon$ provides on the one hand the proper treatment of time-ordered correlation functions, while it also renders the oscillatory integral over the Schwinger parameter $\mathfrak{t}_{L}$ finite due to the convergence factor $e^{-\ep\mathfrak{t}_{L}}$. 

In string theory, the worldsheet is typically considered in Euclidean signature. The analogue of the integral over $\mathfrak{t}_{E}$ in Eq. \eqref{eq:SchwingerProps} in string theory is the integral over the moduli space of Euclidean worldsheet geometries, typically the moduli space of Riemann surfaces $\mathcal{M}_{g,n}$. In a region of the moduli space $\mathcal{M}_{g,n}$ where the Riemann surface develops a long tube, this tube is parametrized by a ``proper time'' $\mathfrak{t}_E$. Such regions often lead to divergences of the amplitudes. For instance for the Veneziano amplitude,\footnote{We will use the calligraphic $\mA$ for closed string amplitudes and italic $A$ for others.}
\be
A_{V}(s,t)=\int_{0}^{1}dx\,x^{-\alpha' s-2}(1-x)^{-\alpha't-2},
\ee
this becomes manifest with the change of variables $x\longleftrightarrow e^{-\mathfrak{t}_{E}}$ near $x\rightarrow 0$. 

\begin{figure} 
    \centering
    \includegraphics[width=0.6\linewidth]{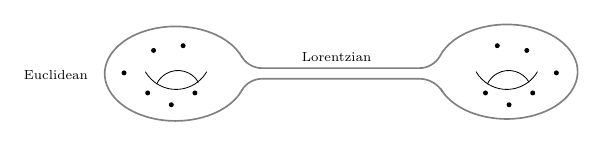}
    \caption{An example of a worldsheet with insertions in closed string theory. The long tube is parametrized by $\frak{t}_E$. For large $\frak{t}_E=T_0\gg 0$, the tube is Wick-rotated to Lorentzian signature.} 
    \label{fig:worldsheet}  
\end{figure}

\begin{figure}
    \centering
    \includegraphics[width=0.6\linewidth]{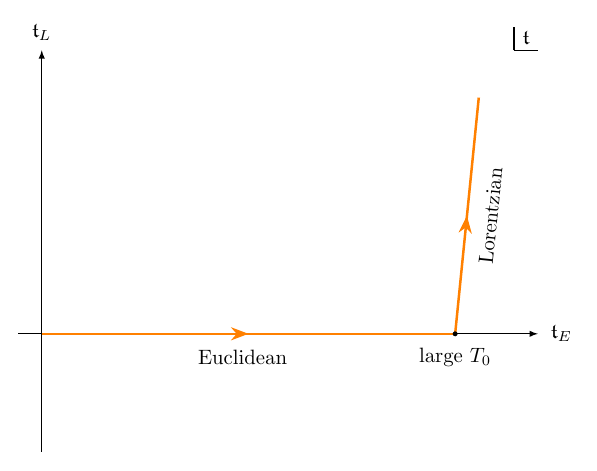}
    \caption{On the $\mathfrak{t}$-plane, the integration contour runs along the real axis to a large value $T_0$. The integration contour then continues in the purely imaginary direction $T_{0}+i\mathfrak{t}_L$ due to the transformation to Lorentzian signature. 
    }
    \label{fig:pathintegralt}
\end{figure}

\begin{figure}
    \centering
    \includegraphics[width=0.6\linewidth]{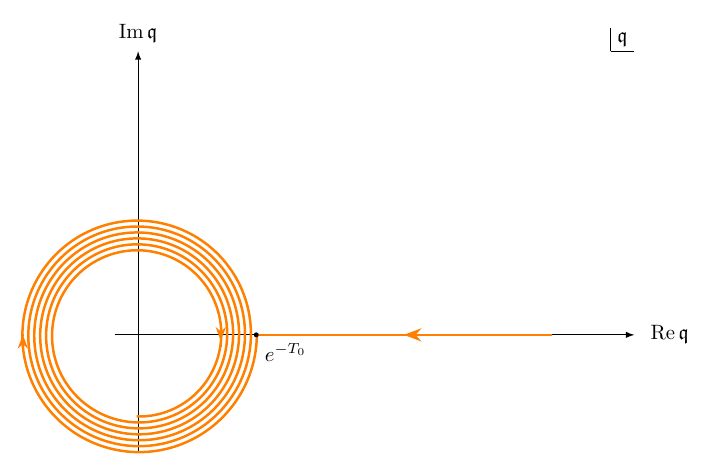}
    \caption{On the $\mathfrak{q}$-plane, the integration contour runs in the decreasing direction along the real axis up to a small value $\mathfrak{q}_{0}=e^{-T_{0}}$. Due to the transformation to Lorentzian signature, the integration contour continues as an infinite spiral around the singular point $\mathfrak{q}=0$ at fixed radius $\mathfrak{q}_{0}$. (The inward spiral is only to visualize that the contour encircles the origin infinitely many times.)}
    \label{fig:pathintegralq}
\end{figure}

To make contact with the $i\ep$-prescription, it is thus natural to apply a Wick rotation to the Lorentzian signature of the worldsheet.
Indeed within string field theory, the vertices are described by a worldsheet with Euclidean signature while the tubes which connect the vertices have Lorentzian signature~\cite{Witten:2013pra, Sen:2024nfd} (see Fig. \ref{fig:worldsheet} for intuition). Thus the Schwinger parametrization becomes a combination of Eq. \eqref{eq:SchwingerProps} and Eq. \eqref{eq:SchwingerPropL}.
\\   
\indent To make this idea explicit, we combine $\mathfrak{t}_E$ and $\mathfrak{t}_L$ to a complex parameter $\mathfrak{t}=\mathfrak{t}_E+i\mathfrak{t}_L$, and apply the Wick rotation from Euclidean to Lorentzian signature at $\mathfrak{t}_E=T_0\gg 0$. The integration can be understood as integrating the proper time over the contour depicted in Fig.~\ref{fig:pathintegralt}. On the $\mft$-plane, the Euclidean contour running along the real axis up to a large proper time $T_{0}$ where the integral may diverge because of the long tube of the degenerate worldsheet. The tube worldsheet is then Wick-rotated to Lorentzian signature, and the integration contour continues vertically along the imaginary Lorentzian time,  $\mft=T_{0}+i\mft_{L}$. 
In the space that is parametrized by $\mathfrak{q}=e^{-\mft}$ (Fig.~\ref{fig:pathintegralq}), the picture above transforms into a contour that moves radially inward along the real axis, and then rotates for infinitely many loops of a small radius $e^{-T_{0}}$ around $\mathfrak{q}=0$.

In string perturbation theory, we similarly apply the Wick rotation in the regions of the moduli space where a long tube develops. See Fig. \ref{fig:worldsheet}. The integrals for the Lorentzian signature are typically convergent since the integrand of the amplitude includes a convergence factor $\mathfrak{t}_L^{-s}$ or $\mathfrak{t}^{-s}$, with $s>1$ taking the role of the convergence factor $e^{-\mathfrak{t}_L\ep}$ in Eq. (\ref{eq:SchwingerPropL}). See for example Eq. \eqref{equa:closedstringintegral} and Eq. \eqref{eq:IntC0C1} below.

More generally, if we implement the Lorentz $i\ep$ in the Feynman propagator, the integration cycle for the amplitude becomes a cycle in the complexification of the moduli space~\cite{Witten:2013pra}, which has been shown to be equivalent to the approach in string field theory~\cite{Sen:2016ubf}. In particular, the new integration contour on the fundamental domain specifies a cut-off proper time $T_0$, which plays an important role in evaluating the one-loop string amplitudes~\cite{Eberhardt:2023xck} and renders the potentially divergent integrals finite. In the following, we apply this to one-loop amplitudes of both closed and open strings.

\section{Evaluating Torus Amplitudes of Closed Strings}
\label{sec:closed1loop}
The evaluation of closed string amplitudes can be illustrated rather explicitly for torus amplitudes of closed strings. This section will mostly consider the one-loop contribution $\mathcal{A}_{1,0}$ to the vacuum energy of bosonic closed oriented strings. The tachyonic mode of the theory renders the perturbative vacuum unstable. We will review the evaluation of $\mathcal{A}_{1,0}$ using both the regularization of the exponential integrals~\cite{Marcus:1988vs, Korpas:2019ava}, and the $i\varepsilon$-prescription~\cite{Eberhardt:2023xck}. Both approaches result in the same numerical value  (\ref{eq:AiepsNum}).
Note this involves an imaginary part, which is naturally interpreted as a decay width. Having discussed this integral, it is relatively straightforward to consider more general amplitudes, such as the two-point amplitude $\mathcal{A}_{1,2}$ in Section \ref{sec:2ptAmp}.
  
 We start by reviewing the integration domain. In Euclidean signature, the closed string propagator is an integral over the proper time $\mft_E$, or imaginary time Schwinger parameter, and the twist angle $x\in [-1/2,1/2]$. These combine to the complex parameter $\tau=x+i\mft_E$, taking values in the Teichm\"uller space for the torus, i.e. the upper-half plane $\mathbb{H}$. We will parametrize $\mathbb{H}$ by $x+iy$ with $x\in \mathbb{R}$ and $y>0$. The natural integration domain for $\tau$ in one-loop amplitudes is the fundamental domain $\mathcal{F}=\mathbb{H}/\rSL(2,\mathbb{Z})$ which parametrizes the complex structures of the Euclidean torus. The canonical choice for $\CF$ is the key-hole fundamental domain,
 \be
\mathcal{F}_\infty=\left\{x+iy\in \mathbb{H}\,|\,x\in[-1/2,1/2],\,y\in[\sqrt{1-x^{2}},\infty)\right\}.
 \ee 
 
Our main example is the vacuum amplitude $\mathcal{A}_{1,0}=\mathcal{A}_{0}$, which up to a prefactor reads~\cite{Polchinski:1985zf, Polchinski:1998rq}\footnote{The prefactor for the vacuum amplitude is $V_{26}/(4(4\pi^2\alpha')^{13})$. Since we restrict to one-loop amplitudes in this paper, we will omit the subscript $g$ in the following from the amplitude $\mA_{1,n}=\mA_{n}$.}
 \be 
\mathcal{A}_0=i\,I_{0},
 \ee 
with the integral $I_{0}$ defined by
\begin{align}
\label{eq:Iclosed}
I_{0}&=\int_{\mF}d\tau\wedge d\bar{\tau}\frac{1}{y^{14}\,\left|\eta(\tau)\right|^{48}},
\end{align}
where $\tau=x+iy$ and $\eta(\tau)$ the Dedekind eta function defined in Eq. (\ref{eq:defeta}). We recall the Fourier expansion,
\be 
\label{intfFourier}
\begin{split}
\frac{1}{\eta(\tau)^{24}}&=\sum_{n=-1}^\infty F(n)\,q^{n}\\
&=q^{-1}+24+324\,q+\dots,
\end{split}
\ee 
with Fourier coefficients $F(n)$ defined as in Eq. (\ref{eq:FC}). The polar term $q^{-1}$ is due to the tachyonic mode in bosonic string theory and diverges for $\tau\to i\infty$. Consequently, the integral  \eqref{eq:Iclosed} is clearly divergent. This is quite a general phenomenon of torus amplitudes. To discuss such integrals collectively, we note that $I_0$ (Eq. \eqref{eq:Iclosed}) and other $n$-point torus amplitudes of closed strings can be written as  
\begin{align}
\label{eq:genIntf}
\mathcal{I}_f=\int_{\mF}\,d\tau\wedge d\bar{\tau}\,y^{-s} f(\tau,\bar{\tau}),
\end{align}
where $f(\tau,\bar{\tau})$ is a non-holomorphic modular form of weight $(2-s,2-s)$. Such integrals have a long history, partly as providing an inner product on the space of cusp forms~\cite{Petersson1950}, its appearance in one-loop string amplitudes~\cite{Lerche:1988np, Marcus:1988vs, Dixon:1990pc, Harvey:1995fq}, $u$-plane integrals~\cite{Moore:1997pc, Malmendier:2008db, Korpas:2017qdo, Korpas:2019ava, Korpas:2019cwg}, and its use for theta lifts~\cite{Borcherds:1996uda}. 

Section \ref{sec:evieps} discusses the evaluation of $I_0$ using the $i\varepsilon$-prescription, and Section \ref{sec:evExpInt} discusses the regularization using exponential integrals. Section \ref{sec:equivalence} demonstrates the equivalence using a contour deformation. Section \ref{sec:2ptAmp} evaluates a two-point function which provides an example with a different integrand $f$.

\subsection{Evaluation Using the $i\varepsilon$-Prescription}  
\label{sec:evieps}
\begin{figure}
    \centering
    \includegraphics[width=\linewidth]{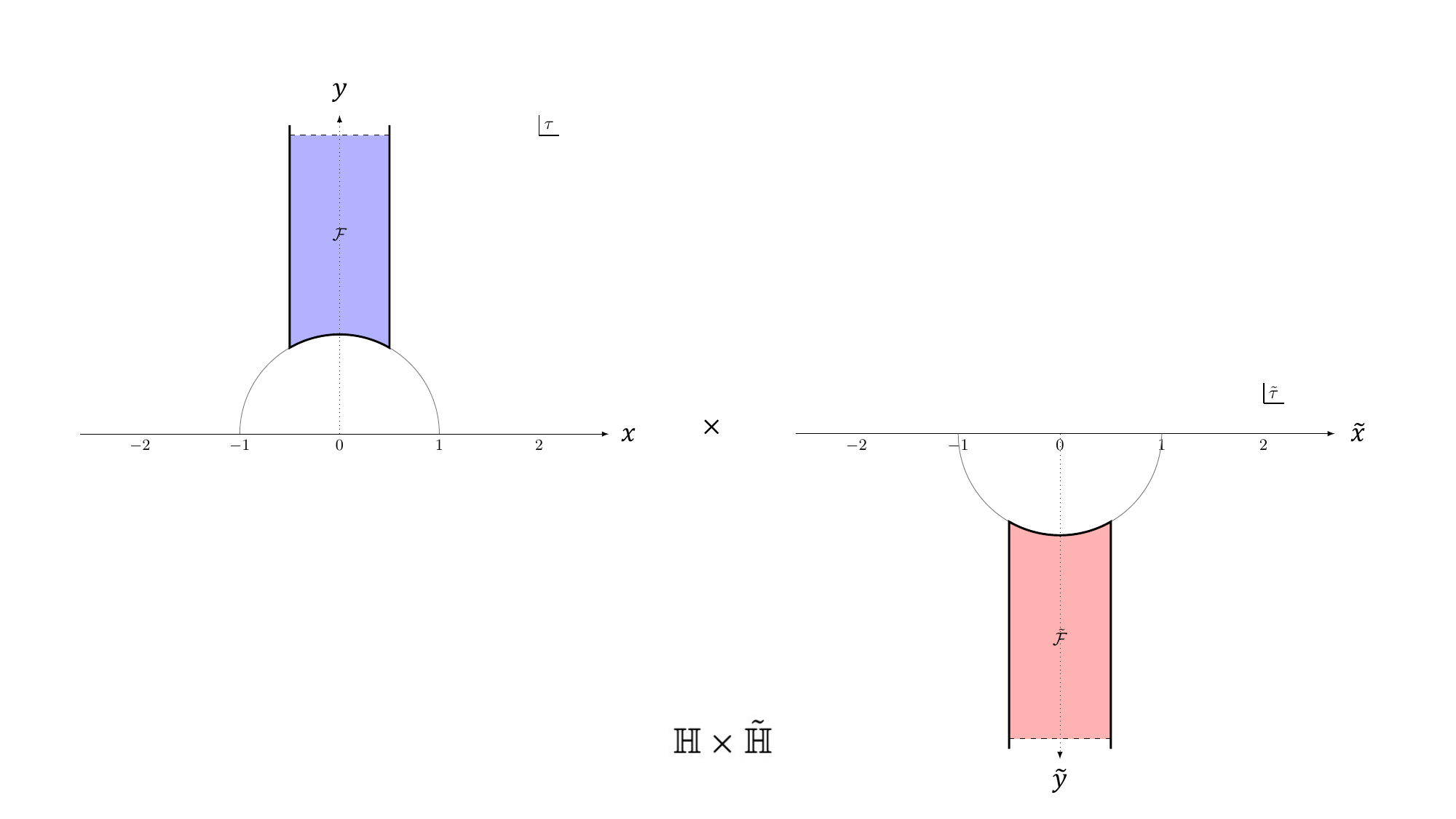}
    \caption{The complexification of the upper-half-plane, $\mH^{\mC}\simeq\mH\times\tilde{\mH}$. The left half displays the upper-half-plane $\mathbb{H}$ parametrized by $\tau=x+iy$. The fundamental domain $\mF$ is displayed in blue. The right half displays the lower-half-plane $\tilde{\mathbb{H}}$ and fundamental domain $\tilde{\mathcal{F}}$ parametrized by $\tilde{\tau}=\tilde x-i\tilde y$. The complex structure of $\tilde{ \mathbb{H}}$ and $\tilde{\mathcal{F}}$ is opposite to that of $\mathbb{H}$ and $\mathcal{F}$.}
    \label{fig:cplxintcycle}
\end{figure}

As explained in the previous subsection, the $i\varepsilon$-prescription provides a finite result by analytic continuation of $\mft_E$ to a complex variable $\mft$ and to integrate $\mft$ over the domain $[iT_0,i\infty)$.
More generally, the $i\varepsilon$-prescription for the evaluation of the closed string amplitude is an integration cycle in the complexification of Teichm\"uller space~\cite{Witten:2013pra}. This is for torus amplitudes the complexification $\mathbb{H}^\mathbb{C}$ of the upper-half-plane $\mathbb{H}$. We have $\mathbb{H}^\mathbb{C}\simeq \mathbb{H} \times \mathbb{\tilde H}$, where the complex structure of $\mathbb{\tilde H}$ is opposite to that of $\mathbb{H}$.\footnote{If $z=x+iy$ is a holomorphic variable for complex structure $J$, a holomorphic variable for the opposite complex structure $J^{\rm opp}=-J$ is $\bar z=x-iy$.} We always orient the planes $(x,y)$ and $(\tilde x,\tilde y)$ such that $x$ and $\tilde x$ are on the horizontal axis, and $y$ and $\tilde y$ are on the vertical axis (see Fig. \ref{fig:cplxintcycle}).

\begin{figure}
	\begin{minipage}[t]{0.45\linewidth}
		\centering
		\includegraphics[width=2.2in]{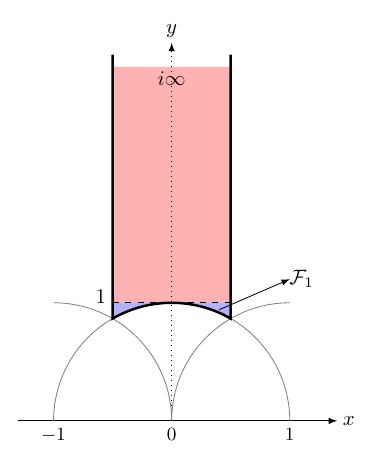}
	\end{minipage}
	\begin{minipage}[t]{0.45\linewidth}
		\centering
		\includegraphics[width=2.2in]{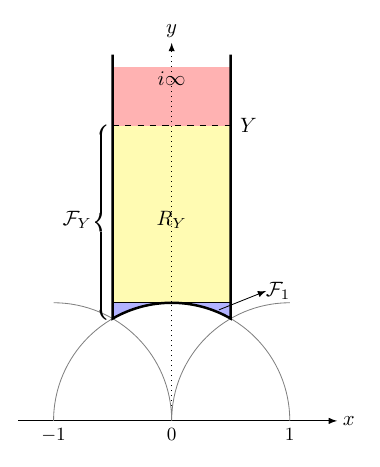}
	\end{minipage}
	\caption{Left panel: The domain $\CF_1$ and the semi-infinite strip $S_1$ (red). Right panel: The semi-infinite strip $S_Y$ (red) illustrates the fundamental domain $R_{\infty}$ ranges up to $y=i\infty$ singularity. The compact rectangle $R_{Y}$ (yellow) and finite keyhole $\mF_{1}$ (blue) regions together illustrates the cut-off fundamental domain $\mF_{Y}$, on which the modular integral will be finite. }
	\label{fig:4.1}
\end{figure}

To explain this in more detail, we introduce the cut-off fundamental domain (see Fig.~\ref{fig:4.1}),
 \be 
 \label{eq:FYdef}
\mathcal{F}_Y=\left\{x+iy\in \mathbb{H}\,|\,x\in[-1/2,1/2],\,y\in \left[\sqrt{1-x^{2}},Y \right]\right\}.
 \ee 

To introduce the new integration cycle, we use the embedding $\iota:\CF_Y\to \mathbb{H}^\mathbb{C}$,
\be
\label{defFY}
\CF_{Y,\mathbb{C}^2}=\left\{(z,\tilde z)\in \mathbb{H}\times  \mathbb{\tilde H} \left|\, z\in \CF_Y\subset \mathbb{C} , \tilde z=\bar z \right.\right\}
\ee
We further introduce the semi-infinite strip $S_{Y}\in \mathbb{H}\times  \mathbb{\tilde H}$, by requiring that the analytic continuation of the real part of $z$, that is $(z+\tilde z)/2$ is in $[-\tfrac{1}{2},\tfrac{1}{2}]_{\mathbb{R}}$ and that the analytic continuation of the imaginary part of $z$, that is $(z-\tilde z)/2$ equals $Y$ plus a purely imaginary number, 
\be
S_{Y}:=\left\{(z,\tilde z)\in \mathbb{H}\times  \mathbb{\tilde H} \,\left| \,\frac{z+\tilde z}{2} \in [-\tfrac{1}{2},\tfrac{1}{2}]_{\mathbb{R}},\,\, \frac{z-\tilde z}{2i}\in Y+ i[0,\infty)_\mathbb{R} \right. \right\}.
\ee 
With $z=x+iy$ and $\tilde z=\tilde{x}-i\tilde{y}$, this is equivalent to
\be
S_{Y}:=\left\{(x,\tilde x,y,\tilde y)\in \mathbb{R}^4\,|\, x+\tilde x\in [-1,1],\, x-\tilde x\in (-\infty,0],\, y+\tilde y=2Y,\, y-\tilde y=0  \right\}.
\ee 
Note that $S_{Y}\in \mathbb{H}\times  \mathbb{\tilde H}$, and not a cycle in the complexification of the torus moduli space $\mathcal{F}\times  \mathcal{\tilde F}$.

The two-dimensional integration cycle $\mathcal{F}_{i\ep}$ reads
 \be
\mathcal{F}_{i\ep}= \left\{ \left.(z,\tilde z)\in \mathbb{H}\times \mathbb{\tilde H}\, \right|   (z,\tilde z)\in \mathcal{F}_{T_0,\mathbb{C}^2}  \cup S_{T_0}  \right\}.
 \ee 
The integration cycle for the vacuum amplitude then runs over the two-dimensional integration cycle $\mathcal{F}_{i\ep}\in \mathbb{H}\times \mathbb{\tilde H}$, such that the closed string integral is given by
\begin{align}
\label{IW}
I_{0}^{i\varepsilon}&=\int_{\mathcal{F}_{i\ep}}d\tau\wedge d{\tilde\tau}  \left( \frac{2i}{\tau-\tilde \tau}\right)^{14}  \frac{1}{\eta(\tau)^{24}\,\eta(-\tilde \tau)^{24}}.
\end{align}
To proceed, we expand $|\eta(\tau)|^{-48}$ as a $(q,\bar q)$-series,
\be
|\eta(\tau)|^{-48}=\sum_{m,n=-1}^{\infty} F(m,n)\,q^m\bar q^n=(q\bar{q})^{-1}+\dots~,
\ee
where the Fourier coefficients $F(m,n)$ are expressed as the product of Fourier coefficients $F(m)$ of $1/\eta^{24}$  (\ref{intfFourier}),
\be
F(m,n)=F(m)\,F(n).
\ee 
Then the closed string amplitude, Eq. (\ref{IW}), can be expressed as an infinite sum over $m$ and $n$,
\be 
\label{eq:I0ieps}
I_{0}^{i\varepsilon}=\sum_{m,n=-1}^\infty F(m,n)\,L^{i\varepsilon}_{m,n,14},
\ee 
with  
 \be
 \begin{split}
L^{i\varepsilon}_{m,n,s}&=\int_{\mF_{T_{0}}}d\tau\wedge d\bar{\tau}\,y^{-s}q^{m}\bar{q}^{n}-2i\int_{-\frac{1}{2}}^{\frac{1}{2}}dx\int_{T_{0}}^{T_{0}+i\infty} dy\,y^{-s}q^{m}\bar{q}^{n}\\
&=\int_{\mF_{T_{0}}}d\tau\wedge d\bar{\tau}\,y^{-s}q^{m}\bar{q}^{n}-2i\delta_{m,n}\int_{T_{0}}^{T_{0}+i\infty} dy\,y^{-s}e^{-4\pi m y}.\label{equa:closedstringintegral}
\end{split}
\ee
We identified in the second integral $x=(\tau+\tilde \tau)/2$ and $y=(\tau-\tilde \tau)/2i$. Numerical evaluation then gives~\cite[p. 83]{Eberhardt:2023xck}\footnote{Our convention for listing numerical values is that the last digit given is rounded using the evaluation of the expression at a higher precision.}
\begin{align}
\label{eq:AiepsNum}
\mA_{0}^{i\varepsilon}=i\,I_0^{i\varepsilon}\approx     58 798.14+ 196 620.04\,i.
\end{align}
This numerical value was already determined in Ref.~\cite[Eq. (19)]{Marcus:1988vs}, and more recently in Ref.~\cite{Korpas:2019ava} using regularization of the integral over the fundamental domain. The exact value of the imaginary part is given in Eq. (\ref{eq:ReI0}). It is interesting to compare the first term in Eq. \eqref{eq:I0ieps} to the value in Eq. \eqref{eq:AiepsNum}. One finds 
\be
\label{eq:numLieps}
i\,L^{i\varepsilon}_{-1,-1,14}\approx 60754.25+196620.04\,i,
\ee
The imaginairy part fully matches with Eq. \eqref{eq:AiepsNum}, while the real part differs by about 3\%.

The numerical value of $L^{i\varepsilon}_{-1,-1,14}$ is most quickly reached for $T_0\gtrapprox 1$ on Mathematica. The same precision takes more time with increasing $T_0$, since the values of the real part of the two terms on the rhs of Eq. \eqref{equa:closedstringintegral} quickly diverge for increasing $T_0$, such that there is a large cancellation between the two terms to arrive at the value in Eq. \eqref{eq:numLieps}. For example, we find for the first term for $T_0=1,5$ and 10,
\be
\begin{split}
&T_0=1:\qquad \int_{\mF_{1}}d\tau\wedge d\bar{\tau}\,y^{-14}\, (q\bar{q})^{-1}\approx 26897.31,\\
&T_0=5:\qquad \int_{\mF_{5}}d\tau\wedge d\bar{\tau}\,y^{-14}\, (q\bar{q})^{-1}\approx 6.546 \times 10^{16},\\
&T_0=10:\qquad \int_{\mF_{10}}d\tau\wedge d\bar{\tau}\,y^{-14}\, (q\bar{q})^{-1}\approx 6.740\times 10^{39}.
\end{split}
\ee

The following subsection will demonstrate that the infinite sum (\ref{eq:I0ieps}) obtained from the complexified domain is equivalent to the regularization of the integral over the fundamental domain.
\subsection{Evaluation Using Exponential Integrals}
\label{sec:evExpInt}
This section reviews the prescription to regularize $I_0$ using the incomplete Gamma function~\cite{Marcus:1988vs}, or equivalently the generalized exponential integral~\cite{Bringmann2016}. This was adopted in Ref.~\cite{Korpas:2019ava} for the evaluation of path integrals for topological quantum field theories on four-manifolds with $b_2^+=1$ to preserve the topological BRST symmetry in correlation functions. We denote the amplitude evaluated this way by
\be 
\mathcal{A}_0^{\rm r}=i\,I_0^{\rm r}.
\ee 

Expressing the integrand as a $(q,\bar q)$-series, we can express the integral  (\ref{eq:Iclosed}) as an infinite sum of terms of the form
\begin{align}
L_{m,n,s}=\int_{\mF}\,d\tau\wedge d\bar{\tau}\,y^{-s}q^{m}\bar{q}^{n}, \end{align}
the triples $(m,n,s)$ satisfy $m,n\in\mR$ and $(m-n)\in\mZ$, $s\in\mZ/2$. The integral is finite for $m+n>0$, or $m+n=0$, $s>1$. The integrand diverges for $m+n<0$ for ${\rm Im}(\tau)\to \infty$. Many integrals of this type have $n\geq 0$ and $m$ bounded below (or vice versa). The standard regularization is to first integrate over $x$ and then integrate over $y$~\cite{Lerche:1988np, Dixon:1990pc, Harvey:1995fq,  Borcherds:1996uda} .

Since $\mF$ is non-compact, the integrand may diverge for $y\rightarrow\infty$ for $m-n<0$, resulting in an improper integral. Nevertheless, one can renormalize and regularize the integral by taking limiting value of integrals over compact domains.
Provided the limit exists, $L_{m,n,s}$ can be defined as
\begin{align}
    L_{m,n,s}=\lim_{Y\rightarrow\infty}L_{m,n,s}(Y),
\end{align}
with
\be 
L_{m,n,s}(Y)=\int_{\mF_{Y}}\,d\tau\wedge d\bar{\tau}\,y^{-s}q^{m}\bar{q}^{n},
\ee 
where $\CF_Y$ is defined in Eq. (\ref{eq:FYdef}). Furthermore, $\mF_{Y}$ can be split into $\mF_{1}$ and a compact rectangle $R_{Y}$ 
\begin{equation}
R_{Y}=\left\{x\in \left[-\tfrac{1}{2},\,\tfrac{1}{2}\right],y\in[1,Y]\right\}
\end{equation}
as shown in the right panel in Fig.~\ref{fig:4.1}. For $Y\to \infty$, we have the semi-infinite strip $R_\infty$.
The integral over $\CF_Y$ thus splits as
\be
\label{eq:LmnsY}
    L_{m,n,s}(Y)=\int_{\mF_{1}}\,d\tau\wedge d\bar{\tau}\,y^{-s}q^{m}\bar{q}^{n}-2i\int_{-\frac{1}{2}}^{\frac{1}{2}}\int_{1}^{Y}\,dx\wedge dy\, y^{-s}q^{m}\bar{q}^{n}.
\ee    
For any triple $m,n,s$, the first term on the rhs is finite and independent of $Y$, since the integration domain is compact and there is no improper singularity in this domain. The $x$-integral in the second term on the rhs evaluates to $\delta_{m,n}$, such that the integral evaluates to the difference of two generalized exponential integrals $E_s(z)$, defined as,
\begin{align}
\label{DefExpI}
    E_{s}(z)=\left\{\begin{array}{cc}
        z^{s-1}\int_{z}^{\infty}e^{-t}t^{-s}dt, & \textrm{for }z\in\mC^{*}, \\
        \frac{1}{s-1}, & \textrm{for }z=0,\ s\neq 1, \\
        0,&\textrm{for }z=0,\ s=1,
    \end{array}\right.%\label{generalizedexp}
\end{align}
where for non-integer $s$, we fix the branch of $t^{-s}$ by specifying that the argument of any complex number $z\in\mC^{*}$ is in the domain $(-\pi,\pi]$. Appendix~\ref{App:C} summarizes various aspects of $E_{s}(z)$. We thus express $L_{m,n,s}(Y)$ as
\be
\label{eq:LmnsEs}
L_{m,n,s}(Y)=\int_{\mF_{1}}\,d\tau\wedge d\bar{\tau}\,y^{-s}q^{m}\bar{q}^{n}-2i\delta_{m,n}(E_s(4\pi m)-Y^{1-s} E_s(4\pi m Y)).
\ee
This expression demonstrates that the limit $Y\to \infty$ converges for $m+n>0$, 
\begin{align}
    L_{m,n,s}&=\lim_{Y\rightarrow\infty} L_{m,n,s}(Y).
\end{align}
On the other hand, the integral is generally divergent for $m+n\leq 0$ since $E_s(4\pi m Y)$ diverges for $m<0$. To treat these cases, the regularization $L^{\mathrm r}_{m,n,s}$ of $L_{m,n,s}$ is {\it defined} by subtracting the term with $E_s(4\pi m Y)$ ~\cite{Korpas:2019ava}. The regularized integral $L^{\textrm{r}}_{m,n,s}$ is thus given by
\begin{align}
    \label{equa:regularizedintegral}
    L^{\textrm{r}}_{m,n,s}
    &=\int_{\mF_{1}}\,d\tau\wedge d\bar{\tau}\,y^{-s}q^{m}\bar{q}^{n}-2i\delta_{m,n}E_{s}(4\pi m),
\end{align}
 For $z\in\mR^{-}$ and $s\geq 1$, the integration contour is deformed to the lower half-plane, and ${\rm Im}(E_{s}(z))$ is defined as ${\rm Im}(E_{s}(z))=\frac{\pi(-z)^{s-1}}{\Gamma(s)}$  (\ref{eq:ImEs}). The regularized integral $I^{\rm r}_0$ is then defined as
\be 
\label{eq:I0r}
I_{0}^{\rm r}=\sum_{m,n=-1}^\infty F(m,n)\,L^{\rm r}_{m,n,14}.
\ee 
While this expression appears different from Eq. (\ref{equa:closedstringintegral}), numerical evaluation of this integral gives for $\mathcal{A}_0^{\rm r}$ the same value as Eq. \eqref{eq:AiepsNum}~\cite{Korpas:2019ava}. We will prove their equivalence in the next subsection.

For the integral $\mathcal{I}_f$ \eqref{eq:genIntf} with an integrand a generic non-holomorphic modular form $f$ of half integral weight $(w,w)=(2-s,2-s)$ and with Fourier coefficients $F(m,n)$, $\mathcal{I}_f$ evaluates to Eq. \eqref{eq:I0r} with $14$ replaced by $2-w$. Its real part reads
\begin{empheq}[innerbox=\colorbox{gray!30}]{equation}
{\rm Re}\left[\mathcal{I}_f\right]=\frac{(2\pi)^{2-w}}{\Gamma(2-w)}\sum_{ m=n<0}F(n,n)\,(-2n)^{1-w},
\end{empheq}
which corresponds to the imaginary part of the amplitudes.
\subsection{Proof of the Equivalence}
\label{sec:equivalence}
To demonstrate the equivalence of $\mA_0^{i\varepsilon}$ and $\mA_0^{\rm r}$, we aim to show the identity
\be
\label{Lieps=Lr}
L^{i\varepsilon}_{m,n,s}=L^{\rm r}_{m,n,s},
\ee
for general $m,n$ with $m-n\in \mathbb{Z}$ and $s>1$. This identity will follow using a contour deformation argument. To this end, we express $L_{m,n,s}$ as
\be 
L_{m,n,s}=\int_{\mF_{T_0}}\,d\tau\wedge d\bar{\tau}\,y^{-s}q^{m}\bar{q}^{n}-2i\delta_{m,n}\, T_0^{1-s}\,E_{s}(4\pi m T_0).
\ee

\begin{figure}
	\begin{minipage}[t]{0.45\linewidth}
		\centering
		\includegraphics[width=2.2in]{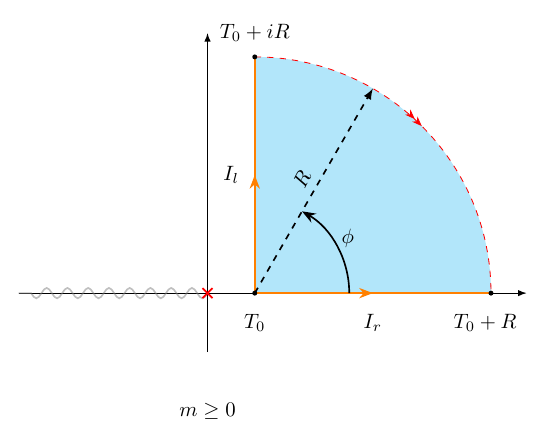}
	\end{minipage}
	\begin{minipage}[t]{0.45\linewidth}
		\centering
		\includegraphics[width=2.2in]{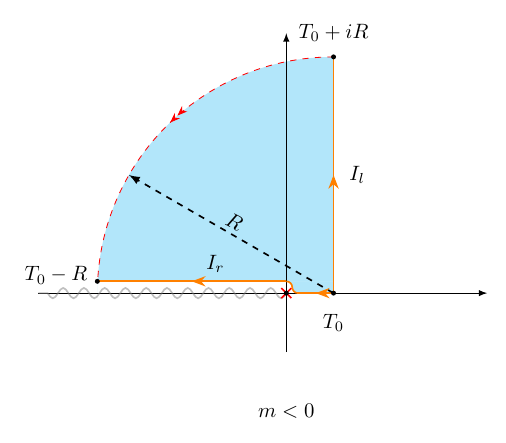}
	\end{minipage}
	\caption{We choose the branch cut along the non-positive real axis $\mathbb{R}^{\leq 0}$, which starts from a singularity of the integrand of order $s-1$ at $0$. In the left panel, $m\geq 0$, and there is no singularity in the region (cyan) swept by the deformation contour (red dashed arc). In the right panel, $m<0$, then the top end $T_{0}+iR$ of the integration domain is deformed to the negative real infinity, so that we should deform the integration domain from $\mathbb{R}^{\leq 0}$ to $\mathbb{R}^{\leq 0}+i\ep$ (we have chosen the branch above the real axis) to circumvent the singularity. It is clear that the contour of $J(m,R)$ around the cyan region does not surround a singularity. Moreover, since the contribution $J_{\phi}(m,R)$ is shown to be vanishing in the cyan region, the equivalence between these two integrations is evident.}
	\label{fig:branchcut}
\end{figure}

Subtracting from the left and right hand side of Eq. (\ref{Lieps=Lr}) the integral over $\CF_{T_0}$, the required identity reduces to
\begin{empheq}[innerbox=\colorbox{gray!30}]{equation}
\label{redLieps=Lr}
\int_{T_{0}}^{T_{0}+i\infty} dy\,y^{-s}\,e^{-4\pi m y}=T_0^{1-s}\,E_s(4\pi m T_0).
\end{empheq}
For the right hand side, we have 
\be
T_0^{1-s}\,E_s(4\pi m T_0)=(4\pi m)^{s-1}\int_{4\pi m T_0}^\infty dt\,t^{-s}\,e^{-t}.
\ee 
For $m<0$ and $s\geq 1$, the integration contour of the rhs is deformed infinitesimally into the lower half plane. The desired identity  (\ref{redLieps=Lr}) can then be written as the equality of the limits ${R\to \infty}$ of the integrals $J_l(m,R)$ and $J_r(m,R)$, defined as
\be 
J_l(m,R)=\int_{T_{0}}^{T_{0}+iR} dy\,y^{-s}\,e^{-4\pi m y},\qquad J_r(m,R)=\int_{T_0}^{T_0+{\rm sgn}(m)R} dy\,y^{-s}\,e^{- 4\pi m y},
\ee 
where ${\rm sgn}(m)=1$ for $m\geq 0$, and ${\rm sgn}(m)=-1$ for $m<0$. 
We can demonstrate this by considering the $R\to \infty$ limit of the contour integral around a quadrant of the circle centered at $y=T_0$,
\be 
J(m,R)=J_l(m,R)-J_r(m,R)+J_\phi(m,R),
\ee 
with 
\be 
\begin{split} 
&m<0:\qquad J_\phi(m,R)=\int_{\pi/2}^{\pi}  dy(\phi)\,y(\phi)^{-s}\,e^{-4\pi m\,y(\phi)}, \\
&m\geq 0:\qquad J_\phi(m,R)=\int_{\pi/2}^{0}  dy(\phi)\,y(\phi)^{-s}\,e^{-4\pi m\,y(\phi)}. 
\end{split}
\ee 
where $y(\phi)=T_0+R\,e^{i\phi}$. The contours are displayed in Fig. \ref{fig:branchcut}. For $m<0$, the contour passes above the branch point and cut as in the figure. The contour integral $J(m,R)$ then vanishes for all values of $m\in \mathbb{R}$, since there are no singularities contained within the contour. 

On the other hand, we can show that the limit $R\to \infty$ of $J_\phi(m,R)$ vanishes as follows. For $m<0$, we can bound $J_\phi(m,R)$ as
\be
\begin{split} 
\left| J_\phi(m,R)\right|&\leq e^{-4\pi m T_0} R^{1-s} \int_{\pi/2}^\pi d\phi\, \left|T_0/R+e^{i\phi}\right|^{-s},
\end{split}
\ee 
%\be
%\begin{split} 
%\left| J_\phi(m,R)\right|&=  e^{-4\pi m T_0} R^{1-s} \left|\int_{\pi/2}^\pi d\phi\, (T_0/R+e^{i\phi})^{-s}\,e^{-4\pi m R e^{i\phi}}\right|\\
%&\leq e^{-4\pi m T_0} R^{1-s} \int_{\pi/2}^\pi d\phi\, \left|T_0/R+e^{i\phi}\right|^{-s}\,e^{-4\pi m R \cos(\phi)}\\
%&\leq e^{-4\pi m T_0} R^{1-s} \int_{\pi/2}^\pi d\phi\, \left|T_0/R+e^{i\phi}\right|^{-s},
%\end{split}
%\ee 
such that for $s>1$,
\be
\lim_{R\to \infty}  J_\phi(m,R)=0.
\ee 
We arrive thus at the equality of limits,
\be
\lim_{R\to \infty} J_l(m,R)= \lim_{R\to \infty} J _r(m,R),
\ee 
which proves the identity  (\ref{Lieps=Lr}). Therefore, we conclude that when $s>1$ (which is valid for string amplitudes), the integration given by implementing the Lorentzian $i\ep$, as a complexification of the associated Euclidean one, is equivalent to the integration given by regularizing the fundamental domain.

Using the formulas of Appendix~\ref{App:C}, we have
\begin{align}
    {\rm Im}\,[E_{s}(x)]=T_{0}^{1-s}\,{\rm Im}\,[E_{s}(T_{0} x)]
\end{align}
when $x<0$. Thus, the imaginary part of the equality  $\eqref{redLieps=Lr}$ for $m<0$ can be rewritten as
\begin{equation}
{\rm Im}\,\left[\int_{T_{0}}^{T_{0}+i\infty}\,dy\,y^{-s}e^{-4\pi my}\right]={\rm Im}[E_{s}(4\pi m)]=\pi\frac{(-4\pi m)^{s-1}}{\Gamma(s)}.
\end{equation}
We stress that the imaginary part of the Lorentzian integral is independent of the choice of $T_{0}$. This feature will play an important role in evaluating the open string amplitudes.

In particular, inserting Eq. \eqref{redLieps=Lr} into Eq. \eqref{equa:closedstringintegral}, the bosonic closed string amplitude is evaluated as
\be
\begin{split}
&\mA_{0}^{i\varepsilon}=\mA_{0}^{\rm r}\\
& =i\sum_{m,n=-1}^{\infty}F(m,n)\left(\int_{\mF_{T_{0}}}d\tau\wedge d\bar{\tau}\,y^{-14}q^{m}\bar{q}^{n}-2i\delta_{m,n}T_{0}^{-13}E_{14}(4\pi mT_{0})\right),\label{equa:bosonicclosedanswer}
\end{split}
\ee
whose imaginary part has a closed form
\begin{align}
\label{eq:ReI0}
    {\rm Re}[I^{i\ep/{\rm r}}_{0}]=2\,{\rm Im}[E_{14}(-4\pi)]=2^{13}\frac{(2\pi)^{14}}{\Gamma(14)}.
\end{align}
This is simply the imaginary part of Eq. \eqref{eq:AiepsNum}.
\subsection{Two-point Amplitude}
\label{sec:2ptAmp}
As a further illustration, we consider the two-point closed string amplitude $\mA_{2}$ with Mandelstam variable $s_{01}=s=1$. This amplitude was evaluated by~\cite{Marcus:1988vs} using analytic continuation of the integral. 

This amplitude is of the form  \eqref{eq:ClosedStringAmp} with $n=2$ and $z_{01}=-z_{1}=z$. The real part of the amplitude contributes to the mass shift, while the imaginary part contributes to the decay width. The integrand over the configuration space is given by the Green function $G(z,\bar z,\tau,\bar \tau)=2\ln|\vartheta_1(\tau,z)/\eta(\tau)^3|-2\pi\,{\rm Im}(z)^2/y$ ~\cite{Marcus:1988vs},~\cite[Eq. (10)]{Stieberger:2023nol}. With $z=u+\tau v$,
\be
\label{eq:torusclosedInt}
    i\int_{\mathbb{T}^{2}}\,dz\wedge d\bar{z}\,e^{2G(z,\bar{z},\tau,\bar{\tau})}= \frac{2y}{|\eta(\tau)^6|^{2}} \int_0^1 du \int_0^1 dv \, e^{-4\pi y v^2 } |\vartheta_1(\tau,z)|^4.
\ee   
The $u$ integral of the product of four theta series is readily evaluated, and reduces the sum over four integers to a sum over three integers. Using one sum, the $v$ integral can be extended from $[0,1]$ to a Gaussian integral over $\mathbb{R}$. Eq. (\ref{eq:torusclosedInt}) then evaluates to 
\be
 i\int_{\mathbb{T}^{2}}\,dz\wedge d\bar{z}\,e^{2G(z,\bar{z},\tau,\bar{\tau})}=   y^{1/2}\left(\bigg{|}\frac{\vartheta_{3}(2\tau)}{\eta(\tau)^{6}}\bigg{|}^{2}+\bigg{|}\frac{\vartheta_{2}(2\tau)}{\eta(\tau)^{6}}\bigg{|}^{2}\right).
\ee
This is a non-holomorphic modular form for $\rSL(2,\mathbb{Z})$ of weight $(-3,-3)$. Note that the second term of the integrand is the modulus squared of the open string integrand  \eqref{equa:1s2pt}, which fits with the double copy relation~\cite{Kawai:1985xq, Stieberger:2023nol}.
The modular integral $I_{2}$ is thus written as\footnote{Note that this is a combination of planar and non-planar contributions, where the $\vartheta_{2}$-term corresponds to planar integral while the $\vartheta_{3}$-term corresponds to non-planar integral.}
\begin{equation}
\begin{split}
    I_{2}&=\int_{\mF}d\tau\wedge d\bar{\tau}\,y^{-9/2}\left(\bigg{|}\frac{\vartheta_{3}(2\tau)}{\eta(\tau)^{6}}\bigg{|}^{2}+\bigg{|}\frac{\vartheta_{2}(2\tau)}{\eta(\tau)^{6}}\bigg{|}^{2}\right).
    \end{split}
\end{equation}
With the discussion from Section \ref{sec:equivalence}, we find that either the $i\ep$-prescription or the regularization gives for this integral,
\begin{equation}
\begin{split}
&    I^{i\ep}_{2}=I^{\rm r}_{2}=\\
    &\quad \sum_{m,n\in\mathbb{N}\atop \&m,n\in-1/4+\mathbb{N}}F(m,n)\left(\int_{\mF_{T_{0}}}d\tau\wedge d\bar{\tau}\,y^{-9/2}q^{m}\bar{q}^{n}-2i\delta_{m,n}T_{0}^{-7/2}E_{9/2}(4\pi mT_{0})\right).
    \end{split}
\end{equation}
The imaginary part has a closed form
\begin{align}
    {\rm Re}[I^{i\ep}_{2}]=2\,{\rm Im}[E_{9/2}(-\pi)]=2\frac{\pi^{9/2}}{\Gamma(9/2)}=\frac{32\pi^{4}}{105}.
\end{align}
Up to an overall prefactor, the numerical evaluation then gives for the amplitude %({\bf JM: factor 2 to be corrected in (3.46-49)})
\begin{align}
 \mA_{2}^{\rm r}\approx 13.923282+29.686580\,i,
  %  \mA_{2}^{\rm r}\approx 27.85+59.37\,i,
\end{align}
This value matches with~\cite[Eq. (18)]{Marcus:1988vs}, where it is also demonstrated that the imaginary part is in agreement with the Optical Theorem for this string amplitude.

\section{Evaluating One-Loop Bosonic Open String Vacuum Amplitudes}
\label{sec:open1loop}
This section considers the one-loop amplitude for the bosonic open string in 26 dimensions, which provides a clear illustration of the various aspects involved. Section \ref{sec:1loopOpen} reviews the vacuum amplitude and the evaluation by Eberhardt-Mizera~\cite{Eberhardt:2023xck}. Section \ref{sec:ImPartAmp} derives an exact expression for the imaginary part. Section \ref{sec:FullIntOpen} carries out the integration directly in terms of exponential integrals. Section \ref{sec:GenInt} generalizes the discussion to the case where the integrand is a generic weakly holomorphic vector-valued modular form of negative weight. This will be useful in Section \ref{sec:TypeIExp}.
\subsection{The $i\ep$-Prescription for the Open Vacuum Amplitude}
\label{sec:1loopOpen}
The one-loop vacuum amplitude $A_0$ of the bosonic open string in 26 dimensions is the sum of two contributions,
\be 
A_{0}=A_{\rm a}+A_{\textrm{M}},
\ee 
the contribution $A_{\rm a}$ from the annulus, and the contribution $A_{\textrm{M}}$ from the M\"obius strip. To parametrize these geometries we consider the complex $z$-plane with identification $z\simeq z+1$ and $z\simeq -\bar z$. The annulus is then obtained by the further identification $z\simeq z +iy$, $y\in \mathbb{R}_{>0}$, while the M\"obius strip corresponds to the identification $z\simeq z+\tfrac{1}{2}+iy$. Up to a common prefactor,\footnote{Compared to Ref.~\cite{Polchinski:1998rq}, Eqs (7.4.3) and (7.4.23), we divide by a factor $iV_{26}/(2(2\pi^2\alpha')^{13})$.} the amplitudes then read for gauge group $\rSO(n)$,
\begin{align}
\label{eq:Aann}
&A_{\textrm{a}}=\frac{n^2}{2^{26}}\int_0^\infty dy\, \frac{1}{\eta(iy)^{24}},\\
\label{eq:AMob}
&A_{\textrm{M}}=\frac{n}{2^{13}}\int_0^\infty dy\, \frac{1}{\vartheta_3(2iy)^{12}\,\eta(2iy)^{12}}.
\end{align}
For the Euclidean theory, the integration runs over $y>0$. The two integrals are ill-defined and have an exponential divergence for $y\to \infty$ due to the tachyon in bosonic string theory. This divergence can be treated using the methods of Section \ref{sec:closed1loop}. It is absent in Type I superstring theory and other supersymmetric theories as a result of the GSO projection. The linear divergence due to the constant terms of the two integrands is cancelled in their sum for $n=2^{13}$ which we assume henceforth. Additionally, the integrals are divergent due to the $y\to 0$ region of the integration domain, which will be discussed below.

\begin{figure}
    \centering
    \includegraphics[width=0.5\linewidth]{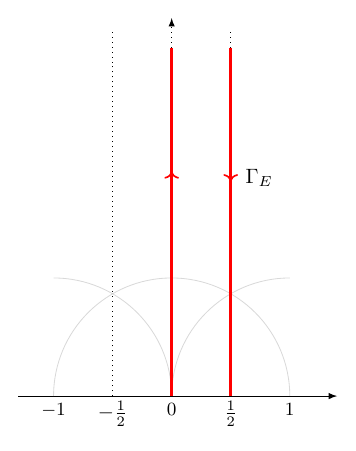}
    \caption{The integration contour $\Gamma_E$ is a union of two vertical lines for the Euclidean open string worldsheet.}
    \label{fig:straightline}
\end{figure}

In analogy with the closed string amplitude, we introduce $\tau=iy$ for the annulus and $\tau=\frac{1}{2}+iy$ for the M\"obius strip. Using the identity  \eqref{eq:thetaeta12eta24},
one finds that the two integrands of Eq. \eqref{eq:Aann} and Eq. \eqref{eq:AMob} become identical as function of $\tau$ for $n=2^{13}$. The open amplitude for Euclidean signature then reads
\be
A_{0,E}=-i \int_{\Gamma_E} d\tau \frac{1}{\eta(\tau)^{24}},
\ee 
with $\Gamma_E$ the contour for $\tau$ displayed in Fig. \ref{fig:straightline}. Note the reversed orientation of the contour at $x=\frac{1}{2}$ relative to $x=0$.

\begin{figure}
	\begin{minipage}[t]{0.45\linewidth}
		\centering
		\includegraphics[width=2.2in]{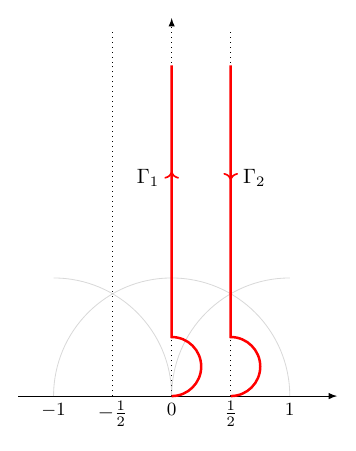}
	\end{minipage}
	\begin{minipage}[t]{0.45\linewidth}
		\centering
		\includegraphics[width=2.2in]{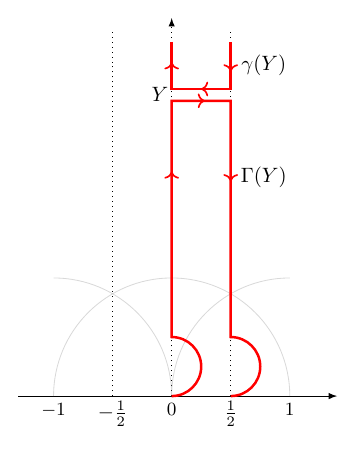}
	\end{minipage}
	\caption{Illustration of the integration contours $\Gamma_1$, $\Gamma_2$, $\Gamma(Y)$ and $\gamma(Y)$. The equivalence of $\Gamma_1\cup \Gamma_2$ and $\Gamma(Y)\cup \gamma(Y)$ is manifest.}
	\label{fig:fullcontour}
\end{figure}

\begin{figure}
    \centering
    \includegraphics[width=0.6\linewidth]{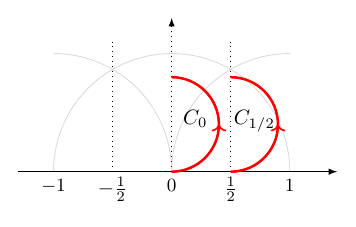}
    \caption{The semi-circles anchored at $\tau=0$ and $\tau=\frac{1}{2}$, respectively. We denote them as $C_{0}$ and $C_{1/2}$. They are equivalent to the Ford semi-circles of fraction $0/1$ and $1/2$.}
    \label{fig:semicircle}
\end{figure}

The discussion on the $i\varepsilon$-prescription suggests an alternative to the contour $\Gamma_E$, namely the contour $\Gamma_1\cup \Gamma_2$ in $\mathbb{H}$, with the two vertical sections near $y=0$ replaced by two semi-circles. We take the radius of the semi-circles to be $\pi/T_0$ for some real parameter $T_0>0$. See Fig. \ref{fig:fullcontour} (left panel).   
We denote the semi-circle anchored at $\tau=0$ by $C_0$, and the one anchored at $\tau=1/2$ by $C_{1/2}$, see Fig. \ref{fig:semicircle}. The semi-circle $C_0$ is parametrized as
\be
    \tau=\frac{2\pi i}{T_{0}+it},
\ee
or equivalently
\be 
\label{eq:xyt}
x(t)=\frac{2\pi t}{T_{0}^{2}+t^{2}},\qquad y(t)=\frac{2\pi T_{0}}{T_{0}^{2}+t^{2}},
\ee 
with $t$ running from $\infty$ to 0. Similarly, we parametrize the integral over $C_{1/2}$ as
\be
\tau=\frac{1}{2}+\frac{\pi i}{2(T_{0}+it)},
\ee 
with $t$ running from $\infty$ to $0$. We thus define the amplitude $A_0^{i\varepsilon}$ following the $i\ep$-prescription as,
\be
A^{i\varepsilon}_{0}=-i \int_{\Gamma_1\cup \Gamma_2} d\tau \frac{1}{\eta(\tau)^{24}}.
\ee 

Note that $\Gamma_1\cup \Gamma_2$ is not a simple contour deformation of $\Gamma_E$. Indeed, if we map $C_0$ to the keyhole fundamental domain $\CF_\infty$, $\tau\to -1/\tau$, it is clear that for $t\to\infty$, $\tau$ does not approach $i\infty$. Yet in a similar spirit to our discussion on the closed string amplitude in Section \ref{sec:closed1loop}, we will prove in Section \ref{sec:RegIntOpen} that the regularized amplitude using exponential integrals gives the identical value as using $\Gamma_1\cup \Gamma_2$.

Eberhardt and Mizera~\cite{Eberhardt:2023xck} decompose the integral over $\Gamma_1\cup \Gamma_2$ by splitting the contour into two contours $\gamma(Y)$ and $\Gamma(Y)$, such that the integral over $\gamma(Y)$ involves the tachyonic divergence. It follows straightforwardly from the residue theorem and explicit computation that the integrals over $\gamma(Y)$ and $\Gamma(Y)$ are separately independent of $Y$. Without loss of generality, we can thus set $Y=T_0$ such that the contour is specified by a single parameter. While the residue theorem also shows that the result is independent of the radii of the semi-circles, this independence is not manifest in the result. We will demonstrate in Section \ref{sec:FullIntOpen} that the integral is independent of this parameter, but that the speed of convergence and numerical accuracy do depend on the choice of $T_0$.

Ref.~\cite{Eberhardt:2023xck} then deforms the contour $\Gamma(T_0)$ to a new contour $\Gamma_\infty$ over arcs of Ford circles analogous to the Circle Method for determining Fourier coefficients of modular forms~\cite{Apostol}. Appendix \ref{app:kloosterman} recalls the expression for the Fourier coefficients. The Circle Method is well used in the microstate counting of black holes, AdS$_3$ gravity and modular bootstrap of 2D CFT, see for example~\cite{Dijkgraaf:2000fq, Denef:2007vg, deBoer:2006vg, Manschot:2007ha, Dabholkar:2014ema, Alday:2019vdr, Iliesiu:2022kny}.
For the $i\varepsilon$-prescription, the crucial difference is that the contour $\Gamma_\infty$, only runs over the arcs of the Ford circles anchored at the fractions in the interval $(0,1/2]$~\cite{Eberhardt:2023xck}. The part of the amplitude corresponding to the contours $\Gamma(T_0)$ and $\Gamma_\infty$, $A_{0,\Gamma}^{i\varepsilon}$ is thus given by~\cite{Eberhardt:2023xck} 
\begin{align}
\label{eq:ASopen0}
    A_{0,\Gamma}^{i\varepsilon}&\equiv-i\int_{\Gamma(T_0)}\frac{d\tau}{\eta(\tau)^{24}} =-i\int_{\Gamma_{\infty}}\frac{d\tau}{\eta(\tau)^{24}}~.
\end{align}    
Using the Circle Method, this is evaluated in terms of an exponential sum, which is reminiscent of a Kloosterman or Ramanujan sum~\cite{Eberhardt:2023xck}. This sum can be deduced from the formula for the Fourier coefficients derived with that method reviewed in \ref{app:kloosterman}. The integral in Eq. (\ref{eq:ASopen0}) is very similar to the integral over Ford circles for the constant term of $\eta^{-24}$ resulting in Eq. \eqref{eq:FConst} specified to $\eta^{-24}$, except that the contour $\Gamma_\infty$ in Eq. (\ref{eq:ASopen0}) runs over the Ford circles anchored at the Farey fractions in the interval $(0,1/2]$ rather than $(0,1]$.

Since $\eta^{-24}$ is a one-dimensional vector-valued modular form with one polar term $q^{-1}$, the result is~\cite{Eberhardt:2023xck}
\begin{align}
\label{eq:ASopen}
    A_{0,\Gamma}^{i\varepsilon}& = -i\frac{(2\pi)^{14}}{\Gamma(14)}\,\mathcal{G}_{14}(-1),
\end{align}
where we introduced the sum $\mathcal{G}_s(n)$, defined as,
\begin{align}
\label{eq:Sopen}
   \mathcal{G}_s(n)\equiv \sum_{c=1}^{\infty}c^{-s}\sum_{-\frac{c}{2} \leq d<0 \atop (d,c)=1} \,e^{\frac{2\pi i n a}{c}}.
\end{align}

The sum is easily implemented and evaluated numerically to high accuracy~\cite{Eberhardt:2023xck}. If we define
\be 
\mathcal{G}_s(n,N)=\sum_{c=1}^{N}c^{-s}\sum_{-\frac{c}{2} \leq d<0 \atop (d,c)=1} \,e^{\frac{2\pi i n a}{c}}
\ee 
The remainder term is bounded by
\be 
|\mathcal{G}_s(n)-\mathcal{G}_s(n,N)|<C\,N^{2-s},
\ee 
for some constant $C$. For example, restricting the sum to $c\leq 10$ gives~\cite{Eberhardt:2023xck},
\be
\frac{(2\pi)^{14}}{\Gamma(14)}\, \mathcal{G}_{14}(-1) \approx -0.001467444355+4.436903\,\times 10^{-6}i.
\ee
The evaluation does become rather slow for increasing $N$ combined with high precision for evaluating the real and imaginary part. We will demonstrate in the next subsection that the real part can be written in the closed form  (\ref{eq:ImAeps}). To our knowledge, no closed form is known for the imaginary part. In Subsection \ref{sec:FullIntOpen}, we express it in a different form in terms of the Fourier coefficients of $\eta(\tau)^{-24}$. 

\subsection{Imaginary Part of $A_{0,\Gamma}^{i\varepsilon}$}
\label{sec:ImPartAmp}
Motivated by unitarity and the optical theorem in string theory~\cite{Okada:1989sd,Marcus:1988vs, Sen:2016gqt, Sen:2016uzq, Eberhardt:2022zay}, this subsection considers the imaginary part of $A_{0,\Gamma}^{i\varepsilon}$  (\ref{eq:ASopen0}), and more generally the real part of $\mathcal{G}_s(n)$. We derive the closed expressions given in Eq. \eqref{eq:Re[G]} and Eq. \eqref{eq:ImAeps}. This is done in two ways, namely using the Kloosterman sum, and using a direct contour evaluation. 

\begin{figure}
    \centering
    \includegraphics[width=0.6\linewidth]{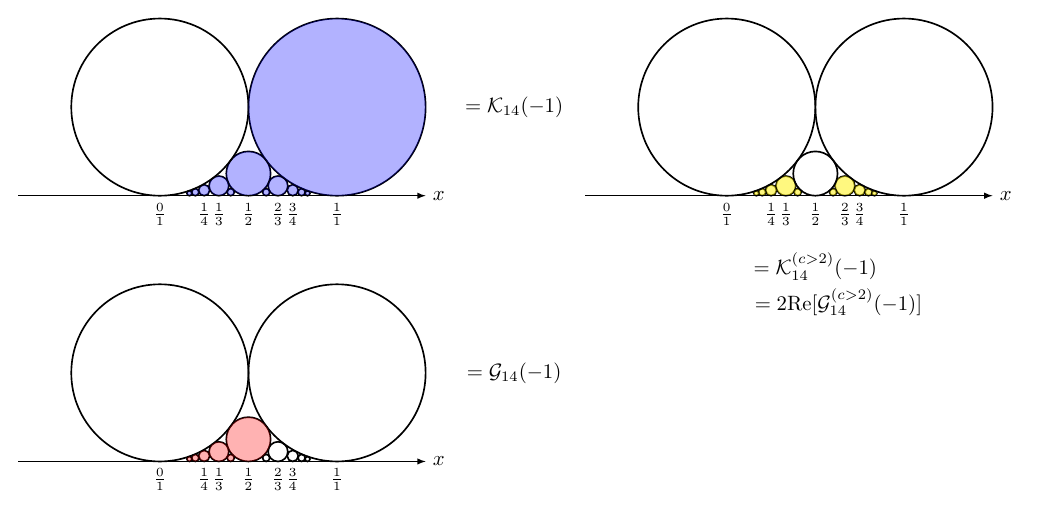}
    \caption{The Ford circles corresponding to the pairs $(c,d)$ contributing to $\mathcal{K}_{14}$ are anchored at the Farey fractions on the interval $(0,1]$; they are displayed in blue in the upper left diagram. The Ford circles corresponding to the pairs $(c,d)$ contributing to $\mathcal{G}_{14}$ are anchored at the Farey fractions on the interval $(0,1/2]$; they are displayed in red in the lower left diagram. The Ford circles corresponding to the pairs $(c,d)$ contributing to $\mathcal{K}^{(c>2)}_{14}$ are displayed in yellow in the right diagram.
    }
    \label{fig:fordcircles}
\end{figure}

We start by relating the real part of $\mathcal{G}_s(m)$ to the sum over Kloosterman sums $\mathcal{K}_s(m,0)$, 
\be
\mathcal{K}_s(m)=\sum_{c=1}^\infty \frac{1}{c^s} K_{c}(m,0),
\ee 
where $K_{c}(m,n)$ is the specialization of the general Kloosterman sum defined in Eq. \eqref{KloosterSum} to $\eta^{-24}$. Since $\eta^{-24}$ has a trivial multiplier system and weight $w=-12$. With the notation of \ref{app:kloosterman}, we have $s=2-w=14$, and $m-\Delta_{\mu}=-1$, $n-\Delta_{\nu}=0$. We thus arrive for $K_c$ at
\begin{align}
    K_{c}(-1)=\sum_{-c< d \leq 0\atop (c,d)=1}e^{-2\pi i\frac{a}{c}}.
\end{align}
Ramanujan's formula gives an exact result for this sum~\cite{apostol1998}:
\begin{align}
    \sum_{c=1}^{\infty}\frac{1}{c^{s}}K_{c}(m,0)=\frac{\sigma_{1-s}(|m|)}{\zeta(s)}.\label{ramanujanformula}
\end{align}
We also introduce the sum over Kloosterman sums for $c>2$,
\be
\label{Kc>2}
\mathcal{K}^{(c>2)}_s(m)=\sum_{c=3}^\infty \frac{1}{c^s} K_c(m,0),
\ee 
such that
\begin{align}
    \mK_{14}(-1)=1-\frac{1}{2^{14}}+\mK^{(c>2)}_{14}(-1,0).
\end{align}
Meanwhile,
\begin{align}
    \mathcal{G}_{14}(-1)&=\frac{1}{2^{14}}e^{\pi i}+\sum_{c>2}\sum_{ -\frac{c}{2} \leq d <0 \atop (c,d)=1} c^{-14}e^{-2\pi i\frac{a}{c}}\\
    &=-\frac{1}{2^{14}}+\mathcal{G}_{14}^{(c>2)}(-1),\n
\end{align}
where we introduced $\mathcal{G}_{s}^{(c>2)}$ similarly to Eq. (\ref{Kc>2}). We then note
\begin{align}
\label{equ:remaindercircles}
    \mK_{14}^{(c>2)}(-1)=2\,{\rm Re}\!\left[\mathcal{G}_{14}^{(c>2)}(-1)\right].
\end{align}
such that we can evaluate the real part of $\mathcal{G}_{s}$ using Eq. \eqref{ramanujanformula},
\begin{align}
    &\mK_{14}^{(c)}(-1)=1-\frac{1}{2^{14}}+2{\rm Re}\!\left[\mathcal{G}_{14}^{(c>2)}(-1)\right]=\frac{1}{\zeta(14)}.
\end{align}    
This then gives 
\be 
\label{equ:realSsum}
{\rm Re}\left[\mathcal{G}_{14}(-1)\right]={\rm Re}\left[\mathcal{G}_{14}^{(c>2)}(-1)\right]-\frac{1}{2^{14}}=\frac{1}{2}\left(\frac{1}{\zeta(14)}-1-\frac{1}{2^{14}}\right).
\ee
From the perspective of the sum over Farey fractions, we can understand this expression as the sum over all Farey fractions minus the one for the Ford circle at $1/1$ and with the Ford circle at $1/2$ circle contributing twice to the real part of $\mathcal{G}_{14}$ (see Fig. \ref{fig:fordcircles}). More generally, we have
\begin{empheq}[innerbox=\colorbox{gray!30}]{equation}
\label{eq:Re[G]}
    {\rm Re}\left[\mathcal{G}_s(n)\right]=\frac{1}{2}\left[\frac{\sigma_{1-s}(|n|)}{\zeta(s)}-1+\frac{(-1)^{n}}{2^{s}}\right].
\end{empheq}
For the imaginary part of the amplitude, we thus arrive at a new expression of the sum $\mG_{s}(n)$ of~\cite{Eberhardt:2023xck},
\be
\label{eq:ImAeps}
{\rm Im}[A_{0,\Gamma}^{i\ep}]=\frac{(2\pi)^{14}}{2\,\Gamma(14)}\left(1+\frac{1}{2^{14}}-\frac{1}{\zeta(14)}\right).
\ee 
In the following, we would like to understand this expression from a more direct calculation. Our discussion here is similar to~\cite{Eberhardt:2022zay}. We aim to evaluate the integral
\be 
\label{eq:intf}
I_{\Gamma(T_0)}=\int_{\Gamma(T_0)}\,d\tau f(\tau),
\ee 
with $f(\tau)=1/\eta(\tau)^{24}$ and Fourier expansion as in Eq. (\ref{intfFourier}). The integration domain $\Gamma(Y)$ is the contour in Fig. \ref{fig:fullcontour}. The real part of this integral then reads
\be 
2\,{\rm Re}[I_{\Gamma(T_0)}]=\int_{\Gamma(T_0)}\,d\tau f(\tau)+\int_{\Gamma(T_0)^{*}}\,d\bar\tau \bar{f}(\bar\tau),
\ee 
with $\Gamma(Y)^{*}$ the complex conjugate of the contour $\Gamma(Y)\in \mathbb{H}$. We explicitly write this as a sum over 10 integrals
\be
\label{eq:GammaInt}
\begin{split}
&2\,{\rm Re}[I_{\Gamma(T_0)}]=\int_{C_0} d\tau\,f(\tau)+\int_{\frac{2\pi }{T_{0}}}^{T_{0}}idy\,f(iy)+\int_{0}^{1/2}dx\, f(x+iT_0)\\
&+\int_{T_{0}}^{\frac{2\pi }{T_{0}}}idy\, f(\tfrac{1}{2}+iy)-\int_{C_{1/2}} d\tau\,f(\tau)\\
&+\int_{C_0} d\bar \tau\,f(-\bar \tau)+\int^{T_{0}}_{\frac{2\pi }{T_{0}}}(-idy)\,f(iy)+\int_{0}^{1/2}dx\, f(-x+iT_0)\\
&+\int_{T_{0}}^{\frac{2\pi }{T_{0}}}(-idy)\, f(\tfrac{1}{2}+iy)-\int_{C_{1/2}} d\bar \tau\,f(-\bar\tau),
\end{split}
\ee 
where the semi-circles $C_0$ and $C_{1/2}$ are displayed in Fig. \ref{fig:semicircle}, and the integration is in both case counter-clockwise. The contributions from the vertical lines cancel exactly (the four integrations over the interval $[\frac{2\pi}{T_{0}},T_{0}]$). The integrals over the interval $[0,\tfrac{1}{2}]$ combine to an integral over $[-\tfrac{1}{2},\tfrac{1}{2}]$, to which only the constant term $F(0)$ of the Fourier series  \eqref{intfFourier} of $f$ contributes. The two integrals along $C_0$ can be combined to an integral over the full circle anchored at $\tau=0$, with $t$ running from $\infty$ to $-\infty$.
Similarly, the two integrals over $C_{1/2}$ can be combined in the same way, but anchoring at $\tau=1/2$.

\begin{figure}
    \centering
    \includegraphics[width=0.6\linewidth]{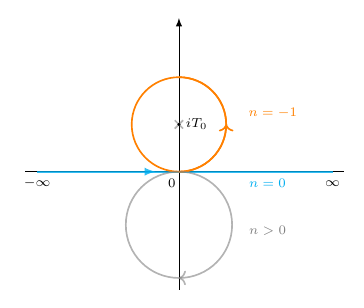}
    \caption{The contour contributes only when it picks up the pole at $t=iT_{0}$ with $n=-1$, all terms with $n\geq 0$ are vanishing after deformation.}
    \label{fig:deformingcontour}
\end{figure}

Combining the above, we thus arrive at 
\be
\begin{split}
2\,{\rm Re}[I_{\Gamma(T_0)}]&= F(0)\\
&-\int_{-\infty}^\infty \frac{2\pi dt}{(T_0+it)^2}\, f\!\left(\frac{2\pi i}{T_0+it}\right)\\
&+\int_{-\infty}^\infty \frac{\pi dt}{2(T_0+it)^2}\, f\!\left(\frac{1}{2}+\frac{\pi i}{2(T_0+it)}\right).\label{equa:intoversemicircle}\\
\end{split}
\ee 
Next we use the identity  (\ref{eq:thetaeta12eta24}) and the $S$-transformations of $\eta$  (\ref{eq:etatrafo}) and $\vartheta_3$  (\ref{eq:theta3trafos})
\begin{equation}
    \eta\left(-\frac{1}{\tau}\right)=(-i\tau)^{1/2}\eta(\tau),\qquad \vartheta_{3}\left(-\frac{1}{\tau}\right)=(-i\tau)^{1/2}\vartheta_{3}(\tau),
\end{equation}
to derive for the modular transformation of $f$, 
\be
\label{ftrafos}
f\left(-\frac{1}{\tau}\right)=\tau^{-12} f(\tau),\qquad f\left(\frac{1}{2}-\frac{1}{\tau}\right)=\left(\frac{\tau}{2}\right)^{-12} f\left(\frac{1}{2}+\frac{\tau}{4}\right).
\ee 
We thus arrive at 
\be 
\begin{split}
2\,{\rm Re}[I_{\Gamma(T_0)}]&= F(0)\\
& -(2\pi)^{13} \int_{-\infty}^\infty \frac{dt}{(T_0+it)^{14}}\,f\!\left(\frac{i(T_0+it)}{2\pi}\right)\\
&+ \frac{\pi^{13}}{2}\int_{-\infty}^\infty \frac{dt}{(T_0+it)^{14}}\,f\!\left(\frac{1}{2}+\frac{i(T_0+it)}{2\pi} \right).
\end{split}
\ee 
The integrals can be evaluated by completing the integration contour to a loop in the upper-half plane for terms in the Fourier expansion with exponent $n<0$, and to a loop in the lower-half plane for the terms with exponent $n\geq 0$. See Figure \ref{fig:deformingcontour}. 
Only the contour in the upper-half plane surrounds the singularity at $t=iT_0$, and thus only the polar term contributes,
\be 
2\,{\rm Re}[I_{\Gamma(T_0)}]=F(0)-\frac{(2\pi)^{14}}{\Gamma(14)}F(-1)-\frac{\pi^{14}}{\Gamma(14)}F(-1).
\ee 
We thus arrive at
\be 
{\rm Im}[A^{i\varepsilon}_{0,\Gamma}]=\frac{(2\pi)^{14}}{2\,\Gamma(14)}\left(F(-1)\left(1+\frac{1}{2^{14}}\right)-\frac{\Gamma(14)\,F(0)}{(2\pi)^{14}} \right),
\ee 
where we can substitute $F(-1)=1$ and $F(0)=24$ from the Fourier series  \eqref{intfFourier}.

This form may look rather different from the imaginary part of Eq. (\ref{eq:ASopen}). The equivalence follows from the exact expression for the constant term  (\ref{FourierRadem}). For $w=-12$, $\Delta=1$ case, we have
\begin{align}
    F(0)&=\sum_{n<0}n^{13}F(n)\left(\frac{(2\pi)^{14}}{\Gamma(14)}\sum_{c=1}^{\infty}\frac{1}{c^{14}}\sum_{-c <d\leq 0 \atop (c,d)=1}e^{2\pi i n\frac{a}{c}}\right).
\end{align}
Since there is only one polar term $F(-1)=1$, this evaluates to
\begin{align}
\label{eq:F0eta24}
    F(0)&=\frac{(2\pi)^{14}}{\Gamma(14)}\sum_{c=1}^{\infty}\frac{1}{c^{14}}\sum_{-c<d\leq 0\atop (c,d)=1}e^{2\pi i \frac{a}{c}}\n\\
    &=\frac{(2\pi)^{14}}{\Gamma(14)}\mathcal{K}_{14}^{(c)}(-1,0)=\frac{(2\pi)^{14}}{\Gamma(14)\zeta(14)}\\
    &=24,\n
\end{align}
where we used
\be 
\zeta(2n)=\frac{|B_{2n}|(2\pi)^{2n}}{2\,(2n)!},
\ee 
with $B_{14}=7/6$. This is in agreement with Eq. \eqref{intfFourier}, and thus confirms the equivalence.

\subsection{Direct Integration for $\Gamma(T_0)$ and $\Pi(T_0)$}
\label{sec:FullIntOpen}
In this section, we will derive an alternative expression for the amplitude $A^{i\varepsilon}_{0,\Gamma}$ to Eq. (\ref{eq:ASopen}) through direct integration. The ingredients for the alternative expression  (\ref{equa:bosonicopenanswer}) are the degeneracies $F(n)$ and generalized exponential integrals $E_s$.

We first treat the integrals over the semi-circles $C_{0}$ and $C_{1/2}$ discussed below Eq. \eqref{equa:intoversemicircle}. Inserting the Fourier expansion for $f$, these read
\begin{align}
\label{eq:IntC0C1}
    I_{C_{0}}&=(2\pi)^{13}\sum_{n\geq -1}F(n)\int_{\infty}^{0}\frac{dt}{(T_{0}+it)^{14}}e^{-(T_{0}+it)n}\n\\
    &=(2\pi)^{13}i\sum_{n\geq -1}F(n)\int_{T_{0}}^{T_{0}+i\infty}e^{-nu}u^{-14}\,du,\\
    I_{C_{1/2}}&=\frac{(4\pi)^{13}}{2}\sum_{n\geq -1}(-1)^{n}F(n)\int^{0}_{\infty}\frac{dt}{(T_{0}+it)^{14}}e^{-\frac{1}{4}n(T_{0}+it)}\n\\
    &=\frac{(4\pi)^{13}}{2}i\sum_{n\geq -1}(-1)^{n}F(n)\int_{T_{0}}^{T_{0}+i\infty}e^{-\frac{n}{4}u}u^{-14}\,du,
\end{align}
where $u\equiv T_{0}+it$. Using the identity  \eqref{redLieps=Lr}, we can express the integral over $u$ in terms of an exponential integral $E_{14}$. We thus arrive at
\begin{align}
    I_{C_{0}}&=(2\pi)^{13}i\sum_{n\geq -1}F(n)\, T_{0}^{-13}\,E_{14}(nT_{0}),\\
    I_{C_{1/2}}&=\frac{(4\pi)^{13}}{2}i\sum_{n\geq -1}(-1)^{n}F(n)\,T_{0}^{-13}\,E_{14}\!\left(\frac{n}{4}T_{0}\right).
\end{align}
The real parts of these expressions indeed match with the discussion in the previous subsection.

Furthermore, the integral over the two vertical lines of $\Gamma(T_{0})$ combine to
\begin{align}
    \int_{\rm vert}\,d\tau\,f(\tau)&=\sum_{n\geq -1}F(n)\int_{\frac{2\pi}{T_{0}}}^{T_{0}}\,idy\,e^{-2\pi ny}+\sum_{n\geq-1}F(n)\int_{T_{0}}^{\frac{2\pi}{T_{0}}}(-1)^{n}\,idy\,e^{-2\pi ny}\n\\
    &=i\sum_{n\textrm{ odd}}F(n)\frac{1}{\pi n}(e^{-2\pi n\tilde{T}_{0}}-e^{-2\pi nT_{0}}),\label{equa:intvert}
\end{align}
where we have defined 
\be 
\tilde{T}_{0}\equiv \frac{2\pi}{T_{0}}.
\ee
The vertical lines thus only contribute to the imaginary part. The integral over the horizontal segment reads
\begin{align}
\int_{\rm hor}\,d\tau\,f(\tau)&=\sum_{n\geq-1}F(n)\int_{0}^{1/2}e^{2\pi inx-2\pi nT_{0}}\,dx\n\\
    &=\frac{1}{2}F(0)+i\sum_{n\textrm{ odd}}F(n)\frac{1}{\pi n}e^{-2\pi nT_{0}}\label{equa:inthor}.
\end{align}
Thus, the full integral is given by\footnote{We set $\frac{\delta_{n,{\rm odd}}}{n}\vert_{n=0}=0$ in this sum. We will always follow this convention throughout the paper.}
\be
\label{eq:EvInt}
\begin{split}
    &I_{\Gamma(T_0)}=\frac{1}{2}F(0)\\
    &\quad +i\sum_{n\in \mathbb{Z}\atop n\geq -1}F(n)\left(\frac{\delta_{n,\ {\rm odd}}}{\pi n}e^{-2\pi n\tilde{T}_{0}}+\tilde{T}_{0}^{13}\,E_{14}(nT_{0})-\frac{(-1)^{n}}{2}(2\tilde{T}_{0})^{13}\,E_{14}\left(\frac{n}{4}T_{0}\right)\right).
\end{split}
\ee
This expression for $I_{\Gamma(T_0)}$ thus takes the form of a generating function of the degeneracies $F(n)$ multiplied by exponential functions of the energy levels $n$. Thus qualitatively the integral $I_{\Gamma(T_0)}$ takes the form of a sum of partition functions with inverse temperatures $2\pi\tilde T_0=4\pi^2/T_0$, $T_0$ and $T_0/4$.

 While each term in the sum over $n$ depends on $T_0$, the integral should be independent of $T_0$ by the residue theorem of complex analysis. We prove this by demonstrating that the derivative vanishes,
\be
\label{IntDer}
\frac{\partial I_{\Gamma(T_0)}}{\partial T_0}=0.
\ee 
To this end, note that Eq. (\ref{ftrafos}) gives
\be
\begin{split}
\sum_{n}\delta_{n,\textrm{ odd}}\, F(n)\,q^n&=\frac{1}{2} \left( f(\tau)-f\left(\frac{1}{2}+\tau\right) \right)\\
&=\frac{1}{2}\left( \tau^{12}f\left(\frac{-1}{\tau}\right)-(2\tau)^{12}f\left(\frac{1}{2}-\frac{1}{4\tau}\right) \right).
\end{split}
\ee 
We have therefore for the first term of the imaginary part of $I_{\Gamma(T_0)}$:
\begin{align}
    &\p_{T_{0}}\left\{\sum_{n}F(n)\frac{\delta_{n,\textrm{ odd}}}{\pi n}e^{-2\pi n\tilde{T}_{0}}\right\}\n\\
    =& \frac{1}{\pi} (\tilde T_0)^2 \sum_n \delta_{n,\textrm{ odd}}\, F(n)\,e^{-2\pi n \tilde T_0}\\
    =&\frac{1}{2\pi}(\tilde{T}_{0})^{2}\sum_{n}F(n)\left((\tilde{T}_{0})^{12}e^{-nT_{0}}-(2\tilde{T}_{0})^{12}(-1)^{n}e^{-nT_{0}/4}\right),\n
\end{align}
while the derivative of last two terms in Eq. \eqref{eq:EvInt} gives
\begin{align}
    \sum_{n}F(n)\left(-(2\pi)^{13}T_{0}^{-14}e^{-nT_{0}}+(-1)^{n}2^{25}\pi^{13}T_{0}^{-14}e^{-nT_{0}/4}\right)~.
\end{align}
Adding up the contributions, we indeed deduce that the derivative vanishes, Eq. (\ref{IntDer}), such that $I_{\Gamma(T_0)}$ is independent of $T_0$.

In particular, the one-loop bosonic open string vacuum amplitude can be expressed as
\begin{empheq}[innerbox=\colorbox{gray!30}]{equation}
    A^{\rm i\varepsilon}_{0,\Gamma}=-iI_{\Gamma(T_0)}=\sum_{n=-1}^\infty C(n,T_0)+i\frac{1}{2}\frac{(2\pi)^{14}}{\Gamma(14)}\left(1+\frac{1}{2^{14}}-\frac{1}{\zeta(14)}\right),\label{equa:bosonicopenanswer}
\end{empheq}
with 
\be 
\begin{split}
C(n,T_0)&=F(n)\left(\frac{\delta_{n,{\rm odd}}}{\pi n}e^{-2\pi n\tilde{T}_{0}}+\tilde{T}_{0}^{13}\,{\rm Re}[E_{14}(nT_{0})] \right. \\
&\quad \left. -\frac{(-1)^{n}}{2}(2\tilde{T}_{0})^{13}\,{\rm Re}[E_{14}\!\left(\frac{n}{4}T_{0}\right)]\right).
\end{split}
\ee
This formula has a closed form for the imaginary part of the vacuum amplitude, and we will demonstrate below that with $T_0$ appropriately chosen, it converges faster than Eq. (\ref{eq:ASopen})~\cite{Eberhardt:2023xck}. On the other hand, the positive and negative contributions in the summand lead to cancellations between exponentially large terms. From the polar term, $n=-1$, we can bound the magnitude of $C(-1,T_0)$ by
\be 
|C(-1,T_0)|< {\rm Max}\left(\frac{3}{\pi} e^{2\pi \tilde T_0}, 3\tilde T_0^{13} E_{14}(-T_0), \frac{3}{2}(2\tilde T_0)^{13} E_{14}(-T_0/4)\right).
\ee 
We can also make an estimate of the largest terms in the summand for $n>0$. From the Hardy-Ramanujan-Rademacher formula Eq. \eqref{FourierRadem}, we can bound the coefficients $F(n)$ by
\be
\label{eq:FnBound}
F(n)<\frac{1}{\sqrt{2}} \frac{e^{4\pi\sqrt{n}}}{n^{27/4}}.
\ee
From a basic saddle point approximation, we find
\be 
|C(n,T_0)|<\frac{1}{\sqrt{2}}\,{\rm Max}\!\left( \frac{1}{\pi \, T_0^{31/4}} e^{T_0}, \frac{1}{(2\pi)^{5/2} T_0^{25/4}} e^{2\pi \tilde T_0}, \frac{2}{(4\pi)^{5/2}T_0^{25/4}}e^{8\pi \tilde T_0} \right).
\ee 
\medskip\\
\noindent
{\it Numerical Evaluation}\\
Having determined $A^{i\varepsilon}_{0,\Gamma}$  (\ref{equa:bosonicopenanswer}), we proceed with its numerical evaluation. The imaginary part is clearly identical to that determined in Eq. \eqref{eq:ImAeps}. We express the real part as 
\be 
{\rm Re}[A^{\rm i\varepsilon}_{0,\Gamma}]=B^{i\ep}_{0,\Gamma}(N, T_0)+R(N+1,T_0),
\ee 
where we have introduced the truncated sum
\be 
\label{eq:ATruncSum}
B^{i\ep}_{0,\Gamma}(N, T_0)\equiv\sum_{n=-1}^N C(n,T_0),
\ee 
and the remainder term
\be 
R(N,T_0)=\sum_{n=N}^\infty C(n,T_0).
\ee 
Even though the implementation of Eq. (\ref{eq:EvInt}) on a computer is a bit more work than Eq. (\ref{eq:ATruncSum}), the numerical evaluation is quite fast depending on the choice of $T_0$. The terms within the brackets decrease exponentially for sufficiently large $n$, such that the convergence is quite fast at least for $T_0$ chosen appropriately.  Table \ref{tab:AGammaNum} presents various values of $B^{i\ep}_{0,\Gamma}(N, T_0)$ for different choices of $N$ and $T_0$.

\renewcommand{\thetable}{4.3}
\begin{center}
\begin{table}[ht!]
\centering
\renewcommand{\arraystretch}{2.2}
\begin{tabular}{|r|r|r|r|}
\hline 
 $T_0$ & $N=-1$ & $N=10$ & $N=75$ \\
\hline 
$\frac{1}{2}\pi$ & $6.23\times 10^{15}$ & $-6.26\times 10^{23}$ & $7.57\times 10^{36}$ \\
\hline 
$2 \pi$ & $1.673\times 10^9$ & $-2.5552\times 10^{10}$ & $4.4369027313$ \\
\hline 
$4 \pi$ & $-4.08\times 10^{6}$ & $4.3339$ & $4.4369027312$ \\
\hline 
$10 \pi$ & $-2.21\times 10^9$ & $-1.1953\times 10^9$ & $4.4366480520$ \\
\hline 
\end{tabular}
\label{tab:AGammaNum}
\caption{Table with approximate numerical values for $B_{0,\Gamma}^{i\ep}(N, T_0)\times 10^6$ for various choices of $N$ and $T_0$.}
\end{table}
\end{center}

 We can make a rough estimate for the magnitude of the remainder $R(N,T_0)$. To this end, we split the remainder term in three parts
\be 
\begin{split} 
&R_1(N,T_0)= \sum_{n=N}^\infty F(n)\,\frac{\delta_{n,{\rm odd}}}{\pi n}\,e^{-2\pi n\tilde{T}_{0}},\\
&R_2(N,T_0)= \tilde T_0^{13}\sum_{n=N}^\infty F(n)\, E_{14}(nT_{0}),\\
&R_3(N,T_0)= -\frac{(2\tilde T_0)^{13}}{2}\sum_{n=N}^\infty (-1)^nF(n)\, E_{14}(nT_{0}/4). \\
\end{split}
\ee 
Since all $F(n)>0$, we can bound $R_1$ as follows
\be 
R_1(N,T_0)< \sum_{n=N}^\infty \frac{F(n)}{n}\,e^{-2\pi n\tilde{T}_{0}}.
\ee 
Substituting the bound for $F(n)$  \eqref{eq:FnBound} in the summand, we deduce that the summand in $R_1$ is $<1$ if the power of the $e$ is negative,
\be
4\pi \sqrt{n}-2\pi n \tilde T_0<0, 
\ee 
and thus for $n>4/\tilde T_0^2=T_0^2/\pi^2$. As a result, for $N>4/\tilde T_0^2$, $R_1$ can be approximated by a geometric sum and bounded as
\be 
\label{eq:R1}
\begin{split}
R_1(N,T_0)&<\frac{1}{\sqrt{2} N^{31/4}} \sum_{n=N}^\infty \exp\!\left(4\pi \left(\sqrt{N}+\frac{1}{2\sqrt{N}}(n-N)\right)-2\pi n \tilde T_0 \right)\\
& = \frac{1}{\sqrt{2}\,N^{31/4}} \frac{\exp\!\left( 4\pi \sqrt{N} - 2\pi N \tilde T_0\right)}{1-\exp(\frac{2\pi}{\sqrt{N}}-2\pi \tilde T_0)}, \qquad N>T_0^2/\pi^2.
\end{split}
\ee
We arrive similarly at the bounds for $R_2$ and $R_3$,
\be 
\label{eq:R2R3}
\begin{split}
&R_2(N,T_0)< \frac{\tilde T_0^{14}}{2\sqrt{2}\pi\,N^{31/4}} \frac{\exp\!\left( 4\pi \sqrt{N} - N  T_0\right)}{1-\exp(\frac{2\pi}{\sqrt{N}}- T_0)}, \qquad N>16\pi^2/T_0^2, \\
&R_3(N,T_0)<\frac{(2\tilde T_0)^{14}}{2\sqrt{2}\pi\,N^{31/4}} \frac{\exp\!\left( 4\pi \sqrt{N} - N  T_0/4\right)}{1-\exp(\frac{2\pi}{\sqrt{N}}- T_0/4)},\qquad N>256\pi^2/T_0^2.
\end{split}
\ee 
We deduce that for $N$ sufficiently large, the magnitude of the remainder $R$ is determined by the maximum of $R_1$ and $R_3$. For the four cases in Table \ref{tab:AGammaNum} with sufficiently large $N$, the remainder term $R$ is bounded by
\be 
R(N,T_0)=\left\{\begin{array}{rr} 10^{-5},& \quad (N,T_0)=(11,4\pi), \\ 5\times 10^{-16}, & \quad (N,T_0)=(76,2\pi), \\ 2\times 10^{-71}, & \quad (N,T_0)=(76,4\pi), \\ 6\times 10^{-9}, & \quad (N,T_0)=(76,10\pi), \\   \end{array} \right.
\ee 
which matches well with the Table. We note that for $(N,T_0)=(11,4\pi)$ and $(76,10\pi)$, the value of $N$ is less than the range of $N$ given in Eq. (\ref{eq:R1}) and Eq. (\ref{eq:R2R3}), yet the magnitude of the summands are much smaller than one. The most optimal value of $T_0$ is for which the maximum of the three lower bounds on $N$ for $R_j$ is the smallest. This is for $T_0=4\pi$, for which the bounds for $R_1$ and $R_3$ also coincide, $N>16$.

\medskip
\noindent
{\it Regularized Integral over $\Pi(T_0)$}\\
\label{sec:RegIntOpen}
\noindent We can also consider the amplitude $\mathcal{A}_{0,\Pi}=iI_{\Pi(T_0)}$ with a slightly different contour, namely $\Pi(T_0)$ where the semi-circles of $\Gamma(T_0)$ are replaced with straight vertical lines $y\in (0,T_0]$. Then the integral reads
\be
\begin{split}
I_{\Pi(T_0)}&=\lim_{\sigma \to 0} \int_{i\sigma}^{iT_0} d\tau \left( f\left(\tau\right)-f\left(\frac{1}{2}+\tau\right)\right)
 + \int_{iT_0}^{iT_0+1/2}\,d\tau\,f(\tau)\\
&=\sum_{n\geq -1} F(n) \int_{1/T_0}^{\infty} idy\,\left( y^{-14}\,e^{-2\pi n y}-4(y/2)^{-14}(-1)^n e^{-\pi n y/2} \right)\\
&+ \int_{iT_0}^{iT_0+1/2}\,d\tau\,f(\tau).
\end{split}
\ee
The integral over $y$ is now divergent due to the term with $n=-1$ on the second line. With the regularization using the exponential integral, the regularized integral $I^{\rm r}_{\Pi(T_0)}$ becomes
\be 
\begin{split}
I^{\rm r}_{\Pi(T_0)}&=\frac{1}{2}F(0)\\
&+i\sum_n F(n)\left[ \frac{\delta_{n,\ {\rm odd}}}{\pi n}e^{-2\pi n T_0}+T_0^{13}E_{14}(n\tilde T_0)-\frac{(-1)^n}{2} (2T_0)^{13} E_{14}(n\tilde T_0/4)\right].
\end{split}
\ee 
This is identical to the rhs of Eq. (\ref{eq:EvInt}) with $T_0$ replaced by $\tilde T_0$. Thus the summands are related by the $S$-transformation of $\rSL(2,\mathbb{Z})$. Since we have proven that the integral is in fact a constant as function of $T_0$, the amplitude $\mathcal{A}^{i\varepsilon}_{0,\Gamma}$ and the renormalized amplitude $\mathcal{A}^{\rm r}_{0,\Pi}$ are thus equivalent,
\be 
\mathcal{A}^{i\varepsilon}_{0,\Gamma}=\mathcal{A}^{\rm r}_{0,\Pi}.
\ee 
\medskip\\
\noindent
{\it Regularized Integral over $\gamma(T_0)$}\\
Finally, we can also carry out the integral over $\gamma(T_0)$ in Fig. \ref{fig:fullcontour}. This evaluates to
\be 
\begin{split}
I^{\rm r}_{\gamma(T_0)}&=-\frac{1}{2}F(0)\\
&+i\sum_n F(n)\left[-\frac{\delta_{n,\ {\rm odd}}}{\pi n}e^{-2\pi n T_0}+T_0\,E_{0}(2\pi n T_0)- (-1)^n T_0\, E_{0}(2\pi nT_0)\right]\\
&=-\frac{1}{2}F(0).
\end{split} 
\ee 
\medskip\\
\noindent
{\it Regularized Integral over $\Gamma_1\cup \Gamma_2$}\\
Adding up the contributions, we arrive for the regularized integral $I^{\rm r}_{\Gamma_1\cup \Gamma_2}$ over $\Gamma_1\cup \Gamma_2$ at
\be 
I^{\rm r}_{\Gamma_1\cup \Gamma_2}=i\sum_n F(n)\left[ \frac{\delta_{n,\ {\rm odd}}}{\pi n}e^{-2\pi n T_0}+T_0^{13}E_{14}(n\tilde T_0)-\frac{(-1)^n}{2} (2T_0)^{13} E_{14}(n\tilde T_0/4)\right].
\ee
This thus determines the regularized amplitude $\mathcal{A}^{\rm r}_{0}\equiv \mathcal{A}^{\rm r}_{0,\Gamma}+\mathcal{A}^{\rm r}_{0,\gamma}$. 

We make a comparison with the evaluation using the Circle Method, in a way which allows generalization to more general cases. We first deduce from Eq. (\ref{eq:Sopen}) that the main contribution to the real part of $I_{\Gamma(T_0)}$ from the sum over $c$ is for $c=2$, $-\frac{(2\pi)^{14}}{\Gamma(14)}2^{-14}$. On the other hand, the main contribution from the sum over $c$ for $F(0)$ \eqref{eq:FConst} is for $c=1$, $\frac{(2\pi)^{14}}{\Gamma(14)}$, which demonstrates that the real part of $I_{\Gamma(T_0)}$ is clearly much smaller than $F(0)$. As a result, the imaginary part of the full amplitude (for the contour $\Gamma_E\equiv\Gamma_1\cup \Gamma_2\equiv\Gamma(T_0)\cup \gamma(T_0)$) is to first-order approximation given by $i\,F(0)/2$.
\subsection{Evaluation for a Generic Integrand}
\label{sec:GenInt}
We conclude this section with the expressions for the integrals over $\Gamma(T_0)$ and $\Gamma_\infty$ for a general weakly holomorphic modular form $f$ of weight $w\in \mathbb{Z}/2$ for a congruence subgroup of $\rSL(2,\mathbb{Z})$, and invariant under $\tau\to \tau+1$. By the residue theorem, the integral over $\Gamma_\infty$ equals the integral over $\Gamma(T_0)$  since $f$ does not have singularities in the interior of the upper-half plane.

For the integral over $\Gamma_\infty$, we consider $f$ as an element $g_\mu$ of a vector-valued modular form $g_\nu$, $\nu=0,1,\dots$ with Fourier coefficients $G_\nu(n)$. The integral is then given by the equation for the constant term  (\ref{eq:FConst}), but with the sum over $d$ restricted to $-c/2\leq d<0$. Thus,
\be
\label{eq:GammaInfGen}
\begin{split} 
\int_{\Gamma_\infty} f(\tau)&=2\pi\,i^{-w} \sum_{\delta_\nu<0} \frac{(2\pi|\delta_\nu|)^{1-w}}{\Gamma(2-w)} G_{\nu}(\delta_\nu)\\
&\times \sum_{c=1}^{\infty}c^{-s}\sum_{-\frac{c}{2} \leq d<0 \atop (d,c)=1} M^{-1}(\gamma)^\nu_\mu\,e^{\frac{2\pi i \delta_\nu a}{c}},
\end{split}
\ee 
where $ad\equiv 1\,\textrm{mod}\,c$ and $\gamma=\left(\begin{array}{cc} a & b \\ c & d \end{array} \right)$. For each pair $(c,d)$ in the sum, $a,b\in \mathbb{Z}$ are chosen such that $\gamma\in SL(2,\mathbb{Z})$. The summand is independent of the choice of $a$ and $b$. 
Since $c=1$ does not contribute to the sum, the main contribution is from $c=2$. On the other hand,  the main contribution to the constant term  (\ref{eq:FConst}) is from $c=1$. The constant term is therefore significantly larger than the magnitude of $\int_{\Gamma_\infty} d\tau\, f(\tau)$.

For the integral over $\Gamma(T_0)$, and the regularized integral over $\Pi(T_0)$, we determine three Fourier series related to $f$, namely for $\tau\to i\infty$, $\tau\to 0$ and $\tau\to 1/2$. We introduce $\tau_1$ and $\tau_2$ as local coordinates near $0$ and $1/2$, such that $\tau=-1/\tau_1$ and $\tau=\frac{1}{2}-1/\tau_2$. Together with $f$, we then define natural functions $f_1$, $f_2$ and their Fourier expansions as follows,
\be 
\begin{split}
f(\tau)&=\sum_{n\in \mathbb{Z}\atop n\geq 0} F_\infty(n)\,q^n,\\
f_1(\tau_1)&= (-i\tau_1)^{-w}\, f(-1/\tau_1)\\
&=\sum_{n} F_1(n)\,q_1^n,\\
f_2(\tau_2)&=(-i\tau_2)^{-w}\,f(1/2-1/\tau_2)\\
&=\sum_{n} F_2(n)\,q_2^n.\\
\end{split} 
\ee 
The functions $f_1$ and $f_2$ can be determined from the transformation law of the vector-valued modular form $g_\nu$  (\ref{eq:fmutrafo}). In particular for $f_2$,
\be 
f(1/2-1/\tau_2)=f(\,\gamma(-1/2+\tau_2/4)\,),\qquad \gamma=\left(\begin{array}{cc} 1 & 0 \\ 2 & 1\end{array}\right).
\ee 

The integral over the cycle $\Gamma(T_{0})$ is given in terms of these Fourier coefficients by 
\begin{empheq}[innerbox=\colorbox{gray!30}]{align} 
 \int_{\Gamma(T_{0})}d\tau\,f(\tau)&=\frac{1}{2}F_{\infty}(0)+i\sum_{n} F_{\infty}(n)\frac{\delta_{n,\ \textrm{odd}}}{\pi n}e^{-2\pi n\tilde{T}_{0}} \n\\
    &+i\sum_{\ell=1,2}\sum_{n}(-1)^{\ell-1}F_{\ell}(n)\,\tilde{T}_{0}^{1-w}\,E_{2-w}(nT_{0}),\label{equ:genformula}
\end{empheq}
with $\tilde T_0=2\pi/T_0$. While the individual terms on the rhs depend on $T_0$, the contour integral is not. The proof is similar to the case of the bosonic open string in Section \ref{sec:FullIntOpen}. This provides a general formula for reproducing the one-loop bosonic open string vacuum amplitudes. While this sum looks more intricate than the formula derived using the Circle Method \eqref{eq:GammaInfGen}, it demonstrates better numerical convergence on $\mathsf{Mathematica}$, and is available to be applied to more general evaluation of one-loop string amplitudes.

Accordingly, the general form of the real part of $I^{\rm}_{\Gamma(T_0)}=I^{\rm}_{\Pi(T_0)}$ is
\begin{empheq}[innerbox=\colorbox{gray!30}]{align}
\label{eq:ImGen}
{\rm Re}\left[\int_{\Gamma(T_{0})}d\tau\,f(\tau)\right]=\frac{1}{2}F_{\infty}(0)-\frac{1}{2}\frac{(2\pi)^{2-w}}{\Gamma(2-w)}\sum_{\ell=1,2}\sum_{n<0}(-1)^{\ell-1}F_{\ell}(n)\,(-n)^{1-w}.
\end{empheq}
This equation involves a sum over polar terms similarly to Eq. \eqref{eq:GammaInfGen} above. This illustrates the connection between the Modular Regularization Method and the Circle Method. It would be interesting to derive this real part from  Eq. \eqref{eq:GammaInfGen} using the approach of Eq. \eqref{ramanujanformula}. 

Summarizing, we have evaluated integrals of the type $\int_{\Gamma(T_0)}d\tau\, f$ in two different ways, namely using the Circle Method in Eq. (\ref{eq:GammaInfGen}) as put forward by \cite{Eberhardt:2023xck} and using exponential integrals in Eq. \eqref{equ:genformula}. We have also discussed the numerical convergence of our results.

\section{Examples of Type I Superstring Amplitudes}
\label{sec:TypeIExp}
To further illustrate the techniques, we evaluate explicitly a few other amplitudes of interest in this section. Section \ref{sec:VacTypeI} considers the one-loop contribution to the vacuum amplitude from the Ramond-Ramond sector in Type I superstring theory, while Section \ref{sec:2point} considers a two-point function of Type I superstring theory. Section \ref{sec:np2pt} further illustrates how to evaluate the non-planar two-point contributions. For each case, we derive expressions for the amplitude both using the Modular Regularization Method as in Eq. (\ref{equ:genformula}) and using the Circle Method as in Eq. \eqref{eq:GammaInfGen}. The comparison involves the integral over the regularized modular domain in terms of the generalized exponential integral and the integral over the Rademacher contour in terms of the Dedekind sum (Eq. \eqref{eq:DS}). Agreement between the two methods is expected for physical and mathematical reasons as Section \ref{sec:equivalence} demonstrated, yet the precise agreement provides a non-trivial check for the evaluations, which are sensitive to many details.

\subsection{Vacuum Amplitude}
\label{sec:VacTypeI}
We illustrate in this subsection our technique by evaluating the one-loop contribution to the closed string vacuum amplitude from the Ramond-Ramond sector in Type I superstring theory. While supersymmetry ensures that this contribution is cancelled by the Neveu/Schwarz-Neveu/Schwarz sector, it is a useful setting for us to see how to apply the regularization in a more general situation other than Section \ref{sec:open1loop}. In this case, we find that the integrand is a weakly holomorphic modular form for the congruence subgroup $\Gamma_0(2)\in \rSL(2,\mathbb{Z})$, and transforms as a vector-valued modular form under the full $\rSL(2,\mathbb{Z})$.

For closed strings of Type I superstring theory, we have three geometries to consider with Euler number $\chi=0$ for tadpole cancellation, the annulus, the M\"obius strip and the Klein bottle. Up to an overall factor, the partition functions for the RR sector read Section 7.4 of~\cite{Polchinski:1998rr}\footnote{Compared to Eqs (10.8.4), (10.8.11) and (10.8.18) in Ref.~\cite{Polchinski:1998rr}, we divide by the factor $\pm iV_{10}/(8(2\pi^2\alpha')^5)$.}
\begin{align}
&A_{\rm a}^{\rm I}= \frac{n^2}{2^{10}}\int_0^\infty dy\, \frac{\vartheta_2(iy)^{4}}{\eta(iy)^{12}}, \label{equa:typeIann}\\
&A_{\textrm{M}}^{\rm I}= \mp\frac{n}{2^4}\int_0^\infty dy\,\frac{\vartheta_{2}(2iy)^{4}\,\vartheta_{4}(2iy)^{4}}{\eta(2iy)^{12}\,\vartheta_{3}(2iy)^{4}},\label{equa:typeImob} \\
&A_{\rm K}^{\rm I}=\int_0^\infty dy\frac{\vartheta_{2}(iy)^{4}}{\eta(iy)^{12}}  \label{equa:typeIkb},
\end{align} 
where the minus sign for $A_{\textrm{M}}^{\rm I}$ is for $\rSO$, and the plus sign for $\textrm{Sp}$ gauge group. Crucially we have for the sum $A_{\rm a}^{\rm I}+A_{\textrm{M}}^{\rm I}+A_{\rm K}^{\rm I}$
\be 
\frac{1}{2^{10}}\int_0^\infty dy\, (n\mp 32)^2\,16+ (n\pm 32)^2\, 256\,e^{-2\pi y}+O(e^{-4\pi y}).
\ee 
The leading constant term famously vanishes for $SO(n=32)$, which causes an IR divergence for $n\neq 32$~\cite{Green:1984ed}. We proceed in the following with $SO(n=32)$. The individual terms in the expansion of the integrand can be integrated, however their sum leads to a UV divergence due to the exponential growth of the coefficients. 

We regularize the sum as before in Section \ref{sec:open1loop} using the contour $\Gamma_1\cup \Gamma_2$ of Fig. \ref{fig:fullcontour}. To arrive at that contour, note that one can show using Eq. \eqref{eq:thetaeta12eta24} that the integrand for the M\"obius strip can be expressed as
\be 
\frac{16}{\vartheta_3(2\tau)^8}=\frac{\vartheta_2(\tau+1/2)^4}{\eta(\tau+1/2)^{12}}.
\ee 
With the identification $\tau=iy$, we can thus express the sum again as an integral over $\Gamma_E$ in Fig. \ref{fig:straightline},
\be 
\begin{split}
A^{\rm I}&=A_{\rm a}^{\rm I}+A_{\textrm{M}}^{\rm I}+A_{\rm K}^{\rm I}\\
&=-2i\int_{\Gamma_E} d\tau\,f(\tau),
\end{split}
\ee 
with 
\be 
f(\tau)=\frac{\vartheta_2(\tau)^4}{\eta(\tau)^{12}}.
\ee 
This expression makes it clear that the integral is divergent due to the integration regions near $0$ and $1/2$. We proceed in the following with the evaluation of the regularized integral for the contour $\Gamma(T_0)$. The evaluation is more intricate compared to the bosonic string since the integrand now forms a three-dimensional representation of $\rSL(2,\mathbb{Z})$. This makes the evaluation of the Kloosterman sums more involved.

To apply the general formula  \eqref{equ:genformula} from Section \ref{sec:GenInt}, we introduce three Fourier series related to $f$, namely for the three singularities $\tau\to i\infty$, $\tau\to 0$ and $\tau\to 1/2$, namely $f$, $f_1$ and $f_2$. Their Fourier expansions read,
\begin{itemize}
\item for $\tau\rightarrow i\infty$,
\be 
\begin{split}
f(\tau)&=\sum_{n\in \mathbb{Z}\atop n\geq 0} F_\infty(n)\,q^n\\
&=16+256\,q+O(q^2),
\end{split}
\ee
\item for $\tau=-1/\tau_{1}\rightarrow 0$,
\be
\begin{split}
f_1(\tau_1)&= \tau_1^4\, f(-1/\tau_1)=\frac{\vartheta_4(\tau_1)^4}{\eta(\tau_1)^{12}}=\sum_{n\in \mathbb{Z}/2 \atop n\geq -1/2} F_1(n)\,q_1^n\\
&= q_1^{-1/2}-8+36\,q_1^{1/2}+O(q_1),
\end{split}
\ee
\item and for $\tau=\frac{1}{2}-\frac{1}{\tau_{2}}\rightarrow \frac{1}{2}$ using Eq. \eqref{eq:thetaeta12eta24},
\be
\begin{split}
f_2(\tau_2)&=\tau_2^4\,f(1/2-1/\tau_2)=\frac{2^8}{\vartheta_3(\tau_2/2)^8}=\sum_{n\in\mZ/4\atop n\geq 0} F_2(n)\,q_2^n\\
& = 256-4096\,q_2^{1/4}+36864\,q_2^{1/2}+O(q_{2}^{3/4}).
\end{split} 
\ee 
\end{itemize}
We note that the coefficients $F_2$ and $F_\infty$ are related as
\be 
F_2(n)=2^4\,(-1)^{4n}\,F_\infty(4n).
\ee 
Numerical evaluation of Eq. \eqref{equ:genformula} with $w=-4$, then gives
\be 
\label{eq:TypeINume}
\int_{\Gamma(T_{0})} d\tau\,f(\tau)\approx -0.011576613-0.020705983\,i.
\ee 
The best convergence still follows from the exponential growth of Fourier coefficients and exponential decay of generalized exponential integrals for sufficiently large $n$. The value of $T_{0}$ with the smallest value of $n$ are $T_{0}=2\pi$ and $n_{\rm max}\geq 4$. 

One easily derives from (Eq. \ref{equ:genformula}) the exact expression for the real part in this case,
\be 
\frac{1}{2}F_\infty(0)-F_1(-1/2)\,\frac{\pi^6}{\Gamma(6)} =8-\frac{\pi^6}{120}.
\ee 
 \medskip \\
{\it Circle Method}\\
We proceed by deriving an expression for $\int_{\Gamma(T_0)}d\tau\,f(\tau)$ which involves an exponential sum as in Eq. \eqref{eq:ASopen} for the bosonic string. To this end, we apply the general formula  \eqref{eq:GammaInfGen} and determine the multiplier matrix $M(\gamma)_\mu^\nu$, $\mu,\nu=1,2,3$, of Eq. (\ref{eq:fmutrafo}) for the vector-valued modular form 
\be 
\left( \begin{array}{c} g_1(\tau) \\ g_2(\tau) \\g_3(\tau)\end{array}\right)=\frac{1}{\eta(\tau)^{12}}\left( \begin{array}{c}  \vartheta_2(\tau)^4\\ \vartheta_3(\tau)^4 \\ \vartheta_4(\tau)^4\end{array}\right), 
\ee 
where $g_1(\tau)=f(\tau)$. More precisely the matrix elements of the first row, $M(\gamma)_1^\nu$. Since $f$ is a modular form of weight $-4$ for the congruence subgroup $\Gamma_0(2)$, $M(\gamma)_1^1=1$ for $\gamma\in \Gamma_0(2)$. The other transformations can be determined by expressing $\vartheta_j$ in term of $\eta$-products and using the transformation  \eqref{eq:etatrafo}, or using the representation of $\vartheta_j^4$ as Eisenstein series for the congruence subgroup $\Gamma(2)$. One obtains that $M(\gamma)_1^\nu$ vanishes, except for the following cases
\be
\begin{array}{lr}
M\left( \begin{array}{cc} a & b \\ c & d\end{array}\right)_1^1=1, &\qquad c=0\mod 2,\\
& \\
 M\left( \begin{array}{cc} a & b \\ c & d\end{array}\right)_1^2=-1, & \qquad c=1\mod 2, d=1\mod 2,\\
 & \\
M\left( \begin{array}{cc} a & b \\ c & d\end{array}\right)_1^3=1, & \qquad c=1\mod 2, d=0\mod 2. 
\end{array}
\ee 
Since only $g_2$ and $g_3$ have a polar term, $q^{-1/2}$ for both functions, only the terms with $c$ odd in the equation for a general integrand  \eqref{eq:GammaInfGen} contribute. Moreover, one can combine the expression for both $\nu=2,3$ to $M(\gamma^{-1})_1^\nu=(-1)^a$. The formula  \eqref{eq:GammaInfGen}, thus specializes for this integral to 
\be 
\label{eq:RRCircle}
\int_{\Gamma(T_0)}d\tau\,f(\tau)=\frac{2\,\pi^6}{\Gamma(6)}\sum_{c\,\,{\rm odd}\atop c>0} \frac{1}{c^6} \sum_{- c/2  \leq d<0 \atop (c,d)=1} (-1)^a\,e^{-\pi i a/c},\qquad ad=1\mod c.
\ee 
Numerical evaluation confirms that the expression converges to the numerical value in Eq. \eqref{eq:TypeINume}. 
\subsection{Planar Two-Point Amplitude}
\label{sec:2point}
Next we consider an example of a planar two-point amplitude from the Type I superstring relevant for the one-loop mass renormalization of \cite[Eq. (20)]{Okada:1989sd}. As before, we combine the annulus and M\"obius strip amplitudes with the contour $\Gamma$. We denote the contribution from $\Gamma(T_0)$ to the planar two-point function with Mandelstam variable $s_{0,1}=s$ by\footnote{The integrand of the non-planar two-point amplitude is the same as in Eq. \eqref{eq:A2Gamma}, but with $\vartheta_1$ replaced by $\vartheta_4$. For $s$ not an even integer, the amplitude can be evaluated with a similar integration contour as $\Gamma(T_0)$, but starting from the cusp at 0 and ending at the cusp at 2~\cite{Eberhardt:2023xck}.}
\be 
\label{eq:A2Gamma}
A_{2,\Gamma}^{\rm I}(s)=-i\int_{\Gamma} d\tau \int_0^1 dz \left(\frac{\vartheta_1(\tau,z)}{\eta(\tau)^3} \right)^{2s},
\ee 
where the integrand is in fact the KN factor~\cite{Stieberger:2023nol} as we discussed in the introduction. Using the Circle Method, Equation (4.44) of~\cite{Eberhardt:2023xck} evaluates $I(s)$ for generic $s$. To illustrate the integral over $\Gamma(T_0)$, we will restrict to $s=1$. 
The integral over $z$ can then be evaluated as~\cite{Stieberger:2023nol}
\begin{align}
    \int_{0}^{1}dz\frac{\vartheta_{1}(z,\tau)^{2}}{\eta(\tau)^{6}}&=\frac{\vartheta_{2}(2\tau)}{\eta(\tau)^6}\label{equa:1s2pt}.
\end{align}
We can then proceed by applying the general formula  \eqref{equ:genformula} with $w=-5/2$. To this end, we consider the Fourier expansions:
\begin{itemize}
\item for $\tau\to i\infty$,
\begin{align}
   f(\tau)&=\frac{\vartheta_{2}(2\tau)}{\eta^{6}(\tau)}=\sum_{n\in \mZ\atop n\geq 0}F_{\infty}(n)\,q^{n}  \\
   &=2+12\,q+O(q^{2}), \n
\end{align}
\item for $\tau=-1/\tau_1\to 0$,
\be 
\begin{split} 
    f_{1}(\tau_{1})&=(-i\tau_{1})^{5/2}f\left(-\frac{1}{\tau_{1}}\right)\\
    &=\sum_{n\in\mathbb{N}-1/4\atop \& n\in\mathbb{N}}F_{1}(n)\,q_{1}^{n}\\&=2^{-1/2}\,(q_{1}^{-1/4} -2 +8\,q^{3/4} -12\,q_1+O(q_1^{7/4})),
\end{split} 
\ee 
\item and for $\tau=\frac{1}{2}-\frac{1}{\tau_2}\to \frac{1}{2}$ using  Eq. \eqref{eq:thetaeta12eta24},
\begin{align}
    f_{2}(\tau_{2})&=(-i\tau_{2})^{5/2}f\left(\frac{1}{2}-\frac{1}{\tau_{2}}\right)=2^{5/2}\frac{\vartheta_{4}(\tau_{2}/2)}{\vartheta_{3}(\tau_{2}/2)^{3}\,\eta(\tau_{2}/2)^{3}}\n\\
    &=\sum_{n\in\mathbb{N}/4-1/16}F_{2}(n)\,q_{2}^{n}\\
    &=2^{5/2}\,q_{2}^{-1/16}(1-8\,q_{2}^{1/4}+O(q_{2}^{1/2})).\n
\end{align}
\end{itemize}
The real part of $\int_{\Gamma(T_{0})}d\tau\,f(\tau)$ receives contributions from the polar terms of $f_1$ and $f_2$,
\be 
\begin{split}
    &{\rm Re}\left[\int_{\Gamma(T_{0})}d\tau\,f(\tau)\right]=\frac{1}{2}F_{\infty}(0)\n\\
    &-F_{1}(-1/4)\,\tilde{T}_{0}^{7/2}\,{\rm Im}\!\left[E_{\frac{9}{2}}\!\left(-\frac{1}{4}T_{0}\right)\right]+F_{2}(-1/16)\,\tilde{T}_{0}^{7/2}\,{\rm Im}\!\left[E_{\frac{9}{2}}\!\left(-\frac{1}{16}T_{0}\right)\right],
\end{split}
\ee 
which evaluates to 
\begin{align}
    &\frac{1}{2}F_{\infty}(0)-\frac{1}{2^{7/2}}\frac{\pi^{9/2}}{\Gamma(9/2)}F_{1}(-1/4)+\frac{1}{2^{21/2}}\frac{\pi^{9/2}}{\Gamma(9/2)}F_{2}(-1/16)\n\\
    &=1-\frac{15}{256}\frac{\pi^{9/2}}{\Gamma(9/2)}=1-\frac{\pi^{4}}{112}.
\end{align}
The two terms agree with the literature~\cite[Eq. (20)]{Okada:1989sd} and \cite[Eq. (4.16)]{Eberhardt:2022zay}. Note that our answer matches with Ref.~\cite{Eberhardt:2022zay} only in the sense of taking the forward limit of double pole degeneration of the associated four-point amplitude. Moreover, the numerical evaluation converges to
\begin{align}
\label{eq:2ptvalue}
 \int_{\Gamma(T_0)} d\tau\,f(\tau)\approx   0.130275973+0.003303550\,i.
\end{align}
The choice $T_0=4\pi$ ensures a rather fast convergence, for which this value is attained for $n_{\rm max}\geq 8$. The numerical value matches with the expression~\cite[Eq. (3.31)]{Eberhardt:2023xck} involving Gauss sums.\footnote{Up to an overall factor $\frac{1}{(2\pi)^{2}}$ due to the normalization convention.} It would be interesting to prove that both values are indeed identical.
We note that this value gives rise to a negative imaginary part for $\mathcal{A}^{\rm I}_{2,\Gamma}(1)$, which is different from the other cases we considered. However, including the contribution from the contour $\gamma(T_0)$ makes the imaginary part of the full amplitude positive as expected.
\medskip\\
{\it Circle Method}\\
We proceed to derive an expression for the amplitude in terms of exponential sums using the Circle Method, and verify that this results in the same value. The evaluation is more involved in this case since the integrand $f$ has half-integral weight. We start by considering the 2-dimensional vector-valued modular form for $\rSL$,
\be 
\label{eq:g01}
\left( \begin{array}{c} g_0(\tau) \\ g_1(\tau) \end{array}\right)=\frac{1}{\eta(\tau)^{6}}\left( \begin{array}{c}  \vartheta_3(2\tau)\\ \vartheta_2(2\tau) \end{array}\right). 
\ee 
In the notation of Appendix \ref{app:kloosterman}, we determine the multiplier system $M(\gamma)_\mu^\nu$ by expressing $\vartheta_j$ as $\eta$-products~\cite{Bringmann:2010sd}. Since the $q$-series for $f=g_1$ does not have a polar term, the elements of interest for us are $M(\gamma)_1^0$. The functions $g_j$ are each modular forms for $\Gamma_0(4)$, such that $M(\gamma)_1^0$ vanishes for $\gamma\in \Gamma_0(4)$. Assuming $d=0\mod 4$ for $c$ odd and $c>0$, we derive for the non-vanishing cases 
\be
\begin{split}
&M\left( \tiny{\begin{array}{cc} a & b \\ c & d \end{array}} \right)_1^0=e^{\frac{\pi i}{12}}\,\xi\!\left( \begin{array}{cc} 2a & 2b-a \\ c/2 & \tfrac{1}{2}(d-c/2) \end{array} \right)^{2}\xi\!\left( \begin{array}{cc} a & 2b \\ c/2 & d \end{array} \right)^{-1}\xi\!\left( \begin{array}{cc} a & b \\ c & d\end{array} \right)^{-6},\quad c=2\mod 4,\\
&M\left( \tiny{\begin{array}{cc} a & b \\ c & d \end{array}}\right)_1^0=\frac{1}{\sqrt{2}}\,\xi\!\left( \begin{array}{cc} 4a & b \\ c & d/4 \end{array} \right)^{2}\xi\!\left( \begin{array}{cc} 2a & b \\ c & d/2 \end{array} \right)^{-1} \xi\!\left( \begin{array}{cc} a & b \\ c & d \end{array} \right)^{-6},\qquad \quad c=1\mod 2,
\end{split}
\ee 
where $\xi(\gamma)$ is the multiplier function of the Dedekind eta function  \eqref{eq:varepsgamma}.
To apply this in the general equation  \eqref{eq:GammaInfGen}, we need to ensure that the bottom left entry of the argument of $M$ is positive. To this end, we use $M(\gamma^{-1})=M(-\gamma^{-1})\,e^{-5\pi i /2}$ by Eq. \eqref{eq:Mids}. We thus arrive at
\be 
\label{eq:planarCircle}
\int_{\Gamma_\infty}d\tau\,f(\tau)= 2\pi\frac{(\pi/2)^{7/2}}{\Gamma(9/2)}\,\mathcal{G}_{9/2},
\ee 
where we introduced the sum
\be 
\mathcal{G}_{9/2}=\,e^{-5\pi i/4}\sum_{c=1}^\infty c^{-9/2}\sum_{-\frac{c}{2} \leq d<0\atop (c,d)=1} M\left(\tiny{\begin{array}{cc} -d & * \\ c & -a \end{array}} \right)_1^0\, e^{-\pi i\frac{a}{2c}},\label{equ:circletwopt}
\ee 
with $a\in \mathbb{Z}$ as usual the modular multiplicative inverse of $d$ modulo $c$, $ad=1\mod c$. Using Eq. \eqref{eq:varepsgamma} and standard identities for the Dedekind sum, this can be simplified to
\be 
\mathcal{G}_{9/2}=\sum_{c=1}^\infty c^{-9/2} \sum_{-\frac{c}{2} \leq d<0\atop (c,d)=1} \chi_{d,c},
\ee 
with the summand $\chi_{d,c}$ defined as
\be 
\chi_{d,c}=\left\{ \begin{array}{rr} \omega^{2}_{a+c/2,c}\, \omega^{-1}_{2a,c}\, \omega^{-6}_{a,c},&\qquad  c=2\mod 4, \\ \frac{1}{\sqrt{2}}\,\omega^2_{a,4c}\,\omega^{-1}_{a,2c}\,\omega^{-6}_{a,c}, & c=1\mod 2, \\ 0,& c=0 \mod 4, \end{array}\right.
\ee 
and $\omega_{d,c}$ given in terms of the Dedekind sum $s(d,c)$  (\ref{eq:DS}) by
\be 
\label{eq:omdc}
\omega_{d,c}=e^{\pi i s(d,c)}.
\ee 
Numerical evaluation converges (slowly) to the value  (\ref{eq:2ptvalue}), which confirms the agreement of the integral over the two contours $\Gamma(T_0)$ and $\Gamma_\infty$.

\subsection{Non-Planar Two-Point Amplitude}
\label{sec:np2pt}
The non-planar two-point amplitude reads
\be 
A^{\rm I,np}_{2}(s)=-i\int_0^{i\infty} d\tau \int_0^1 dz \left( \frac{\vartheta_4(\tau,z)}{\eta(\tau)^3}\right)^{2s}
\ee 
Note that
\be 
-i\int_2^{2+i\infty} d\tau \int_0^1 dz \left( \frac{\vartheta_4(\tau,z)}{\eta(\tau)^3}\right)^{2s}=(-1)^s A^{\rm I,np}_{2}(s)
\ee 
such that
\be
A^{\rm I,np}_{2,\Gamma'_E}(s)=-\frac{i}{2}(1-(-1)^s) \int_{\Gamma'_E} d\tau \int_0^1 dz \left( \frac{\vartheta_4(\tau,z)}{\eta(\tau)^3}\right)^{2s}.
\ee 
where $\Gamma'_E$ is the contour running from $0$ to $i\infty$, and then from $2+i\infty$ to $2$. We then introduce $\Gamma'(T_0)$ similarly to $\Gamma(Y)$ in Figure \ref{fig:fullcontour}, but running between the cusps 0 and 2. Following the notation from Section \ref{sec:open1loop}, we denote the amplitude by $A^{\rm I,np}_{2,\Gamma'}(s)$. In the following we will evaluate this amplitude for $s=1$. We evaluate the integrand for this value as
\begin{align}
f(\tau)=\int_{0}^{1}dz\frac{\vartheta_{4}(\tau,z)^{2}}{\eta(\tau)^{6}}=\frac{\vartheta_{3}(2\tau)}{\eta(\tau)^{6}}.
\end{align}
We define the coefficients $F_\infty(n)$, $F_1(n)$ and $F_2(n)$ by the Fourier expansions near the cusps $i\infty$, $0$ and $2$. One obtains
\begin{itemize}
\item for $\tau\to i\infty$,
\be 
f(\tau)=\sum_{n\in \mZ-1/4}F_{\infty}(n)\,q^{n}=q^{-1/4}+8\,q^{3/4}+O(q^{7/4}).
\ee 
\item for $\tau=-1/\tau_1\to 0$,
\be 
\begin{split} 
f_1(\tau_1)&=(-i\tau_1)^{5/2} f(-1/\tau_1)=2^{-1/2}\frac{\vartheta_3(\tau_1/2)}{\eta(\tau_1)^{6}}=
\sum_{n\in\mathbb{N}-1/4 \atop n\in \mathbb{N}} F_1(n)\,q_1^n \\
&=2^{-1/2}(q_{1}^{-1/4}+8\,q_{1}^{3/4}+2+12\,q_{1}+\cdots).
\end{split} 
\ee 
\item for $\tau=2-\frac{1}{\tau_2}\rightarrow 2$, 
\be
 f_{2}(\tau_{2})=(-i\tau_2)^{5/2} f(2-1/\tau_2)=-2^{-1/2}\frac{\vartheta_{3}(\tau_{2}/2)}{\eta^{6}(\tau_{2})}
 =\sum_{n\in\mathbb{N}-1/4 \atop n\in \mathbb{N}} F_{2}(n)\,q_{2}^{n},
\ee
where we used that $\vartheta_{3}(\tau+2)=\vartheta_3(\tau)$, and $\eta(\tau+2)^{6}=e^{\pi i}\eta(\tau)^{6}$. Thus $F_{1}(n)=-F_{2}(n)$.
\end{itemize}

With these coefficients, we evaluate the integral as
\be 
\int_{\Gamma'(T_0)}d\tau\,f(\tau)=\frac{i}{\pi} \sum_{n\in \mathbb{Z}-1/4} \frac{F_\infty(n)}{n}\,e^{-2\pi n \tilde T_0}+2i\,\tilde T_0^{7/2}\sum_{\mathbb{Z}/4} F_1(n)\,E_{9/2}(nT_0).
\ee 
Numerical evaluation gives
\be 
\label{eq:AInpNum}
A^{\rm I,np}_{2,\Gamma'}(s=1)\approx -1.79524856+1.85541126\,i.
\ee 
The exact value of the imaginary part is $\frac{2}{105}\pi^{4}$.
\medskip \\
{\it Circle Method}\\
To evaluate this amplitude using the Circle Method, we determine the elements $M^0_0$ of the multiplier system of the vector-valued modular form in Eq. (\ref{eq:g01}). Assuming that $d$ is divisible by $4$ for $c$ odd, we obtain
\be 
\begin{split}
&M\left( \tiny{\begin{array}{cc} a & b \\ c & d \end{array}} \right)_0^0=\xi\!\left( \begin{array}{cc} a & 2b \\ c/2 & d\end{array} \right)^{5} \xi\!\left( \begin{array}{cc} a & 4b \\ c/4 & d\end{array} \right)^{-2} \xi\!\left( \begin{array}{cc} a & b \\ c & d\end{array} \right)^{-8},\quad c=0\mod 4,\\
&M\left( \tiny{\begin{array}{cc} a & b \\ c & d \end{array}} \right)_0^0=0,\qquad \quad c=2\mod 4,\\
&M\left( \tiny{\begin{array}{cc} a & b \\ c & d \end{array}}\right)_0^0=\frac{1}{\sqrt{2}}\,\xi\!\left( \begin{array}{cc} 4a & b \\ c & d/4 \end{array} \right)^{-2}\xi\!\left( \begin{array}{cc} 2a & b \\ c & d/2 \end{array} \right)^{5} \xi\!\left( \begin{array}{cc} a & b \\ c & d \end{array} \right)^{-8},\qquad \quad c=1\mod 2.
\end{split}
\ee 
Since the contour runs between the cusps 0 and 2, the contour for the Circle Method traverses the Ford circles anchored between $(0,2]$. We denote this contour by $\Gamma'_{\infty}$. With $-2{c} \leq d<0$, the integral over this contour evaluates to
\begin{align}
\label{eq:NPCircleM}
    \int_{\Gamma'_{\infty}}d\tau\,f(\tau)&=\frac{\pi^{9/2}}{2^{5/2}\,\Gamma(9/2)}\,e^{-5\pi i/4}\sum_{c=1}^\infty c^{-9/2}\sum_{-2{c} \leq d<0\atop (c,d)=1} M\left(\tiny{\begin{array}{cc} -d & * \\ c & -a \end{array}} \right)_0^0\, e^{-\pi i\frac{a}{2c}}\n\\
    &=\frac{\pi^{9/2}}{2^{5/2}\,\Gamma(9/2)}\,\mathcal{G}_{9/2}^{\rm np},
\end{align}
where the sum $\mathcal{G}_{9/2}^{\rm np}$ is defined by the second equality. It can be simplified to,
\begin{align}
    \mG_{9/2}^{\rm np}=\sum_{c=1}^{\infty}\frac{1}{c^{9/2}}\sum_{-2c\leq d<0\atop(c,d)=1}\chi_{d,c}^{\rm np},
\end{align}
with
\begin{align}
    \chi_{d,c}^{\rm np}=\left\{ \begin{array}{rr}  \omega^{-2}_{4a,c}\omega^{5}_{2a,c}\omega^{-8}_{a,c}\,e^{\pi i \frac{d}{2c}},&\qquad  c=0\mod 4, \\ 0,& c=2 \mod 4, \\  \frac{1}{\sqrt{2}}\,\omega^{-2}_{a,4c}\,\omega^{5}_{a,2c}\,\omega^{-8}_{a,c}\,e^{\pi i \frac{d}{2c}}, &\qquad c=1\mod 2,\end{array}\right.
\end{align}
with $\omega_{d,c}$ as in Eq. \eqref{eq:omdc}. As expected, numerical evaluation of Eq. \eqref{eq:NPCircleM} converges to the numerical value for the amplitude given in Eq. (\ref{eq:AInpNum}) times $i$.

\section{Discussion and Conclusion}
\label{sec:conc}
We have explored two strategies for the evaluation of one-loop amplitudes in string theory, for both open or closed strings. For the closed strings, we demonstrated that evaluation using exponential integrals is equivalent to the analytic continuation based on the $i\varepsilon$-prescription. For the open strings, the two different strategies are:
\begin{enumerate}
\item Using the integration contour $\Gamma(T_0)\cup \gamma(T_0)$, which is a string-theoretic generalization of the $i\ep$-prescription of quantum field theory. Using this contour, the amplitudes are expressed in terms of Fourier coefficients and generalized exponential integrals. The imaginary part of the amplitude have a general closed form in this way. This approach is equivalent to regularizing for divergent integrals over modular forms in its own sense~\cite{Bringmann2016, Korpas:2019ava}.
\item Using the integration contour $\Gamma_\infty$ of infinitely many Ford circles put forward by Eberhardt and Mizera~\cite{Eberhardt:2023xck}. Applying this contour, the amplitudes are expressed in terms of arithmetic exponential sums, reminiscent of Ramanujan or Kloosterman sums.
\end{enumerate} 
The equivalence of strategy 1. and 2. follows directly from the residue theorem of complex analysis, yet the final expressions are rather different. The numerical values provide a useful consistency check on each approach. We include Table \ref{tab:Disc} with a comparison of the two methods.

We conclude with listing a few subjects to which our techniques can potentially be applied:
\begin{itemize}
    \item \textbf{The double-copy relation at one-loop level}. The KLT relation works well at tree-level~\cite{Kawai:1985xq} and have been exhaustively explored~\cite{Bern:2010ue,Bern:2019prr, Mafra:2018qqe}. While the KLT relation for one-loop is still an open problem. Refs~\cite{Stieberger:2022lss,Stieberger:2023nol} recently proposed that the integrand for the closed string amplitude is the sum of modulus squared of the integrand of the planar and non-planar open strings. The evaluation of such integrals for closed and open strings is discussed for vacuum amplitudes in Sections \ref{sec:evieps} and  \ref{sec:1loopOpen}, and for two-point amplitudes in Sections \ref{sec:2ptAmp} and \ref{sec:2point}.
    It is fascinating to investigate whether the double-copy relation may also extend beyond the integral over configuration space. 
    \item \textbf{Decay widths and the imaginary parts}. The imaginary parts of the one-loop string amplitudes are relevant for the decay widths and the optical theorem of string amplitudes, see~\cite{Tsuchiya:1988va,Okada:1989sd,Tsuchiya:1989ah,Sen:2014dqa,Sen:2014pia,Eberhardt:2023xck} for relevant work. For a generic integrand, the imaginary part of the amplitude is given by the rhs of Eq. \eqref{eq:ImGen}. We would like to understand physical implications of this expression.
    \item \textbf{Low-energy expansion}. We aim to extend our techniques of direct integration to Regge trajectories~\cite{Okada:1989sd,Banerjee:2024ibt} and to higher-point amplitudes. The latter give rise to more complicated integrals over the $z_{jk}$. This is particularly important for the small $\alpha'$-expansion of the superstring theory, which accounts for the low-energy supergravity approximation and has been investigated for a long time, see for example~\cite{Berkovits:2022ivl} and references therein.
    \item \textbf{Relation to the $u$-plane integral}. As mentioned at a few places in this paper, one-loop string amplitudes share many aspects with $u$-plane integrals in topological field theory \cite{Moore:1997pc, Malmendier:2008db, Korpas:2019ava, Korpas:2019cwg}, such that developments for string amplitudes may also find fruitful applications in these related subjects.
\end{itemize}
\begin{center}
\renewcommand{\thetable}{6}
\begin{table}[ht!]
\centering
\renewcommand{\arraystretch}{2.2}
%\resizebox{!}{\linewidth}
{\begin{tabular}{|p{2.8cm}|p{6.2cm}|p{6.2cm}|}
     \hline 
     & {\bf Modular Regularization Method}  & {\bf Circle Method} \\
   \hline {\bf Ingredients} & Exponential integrals; Fourier coefficients & Arithmetic exponential sums; multiplier systems of modular forms\\   
   \hline
{\bf Applicability} & Easily generalized to other integrands & Dependence on multiplier system makes generalization somewhat tedious\\
   \hline
   {\bf Closed form} & Closed form for ${\rm Im}\,\mA$ is manifest & Closed form for ${\rm Im}\,\mA$ not manifest \\
   \hline
   {\bf Parameter control} & Depends on upper bound $N$ and choice of $T_{0}$ & Only depends on upper bound $N$\\
   \hline
   {\bf Computational performance} & For high accuracy, exponential rate of convergence in $N$, and dependence on choice of $T_{0}$ & Polynomial rate of convergence in $N$ for high accuracy\\
   \hline
   \end{tabular}}
   \label{tab:Disc}
      \caption{Comparison between Modular Regularization Method and Circle Method.}
   \end{table}
\end{center}
\section*{Acknowledgements}
We would like to thank Lorenz Eberhardt, Sebastian Mizera, Greg Moore and Ashoke Sen for discussions. ZZW thanks Yi-Xiao Tao for sharing some references on one-loop string amplitudes. The research of JM was supported by the Ambrose Monell Foundation during his research stay at the Institute for Advanced Study, Princeton, 2022-2023. The research of ZZW was supported by the Postgraduate Award GOIPG/2021/162 of the Irish Research Council.
ZZW would like to thank the Institute for Advanced Study, Princeton, and the workshop ``\textit{BPS Dynamics and Quantum Mathematics}'' of INFN Galileo Galilei Institute for Theoretical Physics, for their hospitality during parts of this work. ZZW also thanks the ICTP Trieste for participation in the ``\textit{String Math 2024}'' and ``\textit{School and Workshop on Number Theory and Physics}''. 

\appendix

\section{Modular Forms}
\label{app:ModForms}
\subsection{The Modular Group and Modular Forms}
The modular group $\rSL(2,\mZ)$ is the group of integer matrices with unit determinant
\begin{align}
    \rSL(2,\mZ)=\left\{\begin{pmatrix}
        a&b\\c&d
    \end{pmatrix}\bigg{|}a,b,c,d\in\mZ;\ ad-bc=1\right\},
\end{align}
which generates a transformation on the fundamental domain on $\mH$ as
\begin{align}
\gamma(\tau)=\frac{a\tau+b}{c\tau+d}.
\end{align}
Two  modular forms appear frequently in the main text:
\medskip \\
{\it Dedekind Eta Function}\\
The Dedekind eta function $\eta:\mH\rightarrow\mC$ is defined as
\begin{align}
\label{eq:defeta}
    \eta(\tau)=q^{1/24}\prod_{n=1}^{\infty}(1-q^{n}).
\end{align}
It is a modular form of weight $\frac{1}{2}$. It transforms under $\gamma=\left(\begin{array}{cc} a & b \\ c & d\end{array} \right)\in \rSL(2,\mathbb{Z})$ as~\cite{apostol1998}
\be 
\label{eq:etatrafo}
\eta\left( \frac{a\tau+b}{c\tau+d}\right)=\xi(\gamma)\,(c\tau+d)^{1/2}\,\eta(\tau),
\ee 
with the $\xi(\gamma)$ determined by
\be
\label{eq:varepsgamma}
\xi(\gamma)=\left\{ \begin{array}{rl} e^{\pi i \frac{b}{12}}, & \qquad c=0,d=1, \\ e^{\pi i \left( \frac{a+d}{12\,c}-s(d,c)-\frac{1}{4}\right)},& \qquad c>0,\end{array} \right.
\ee 
where the function $s(d,c)$ is the Dedekind sum,
\be 
\label{eq:DS}
s(d,c)=\sum_{n=1}^{c-1}  \left( \left(\frac{n}{c} \right)\right)\left( \left(\frac{dn}{c} \right)\right),
\ee 
where $((\,)):\mathbb{R}\to \mathbb{R}$ is defined as
\be 
((x))=\left\{ \begin{array}{rl} x-\lfloor x\rfloor -\frac{1}{2}, & \qquad x \notin \mathbb{Z}, \\ 0, & \qquad x\in \mathbb{Z}.\end{array}\right.
\ee 
\medskip \\
{\it Jacobi Theta Functions}\\
The classical Jacobi theta functions $\vartheta_{j}:\mH\times\mC\rightarrow\mC$, $j=1,\cdots,4$ are given by
\begin{align}
    \vartheta_{1}(\tau,z)&=i\sum_{r\in\mZ+1/2}(-1)^{r-\frac{1}{2}}q^{\frac{r^{2}}{2}}e^{2\pi irz},\\
    \vartheta_{2}(\tau,z)&=\sum_{r\in\mZ+1/2}q^{\frac{r^{2}}{2}}e^{2\pi irz},\\
    \vartheta_{3}(\tau,z)&=\sum_{r\in\mZ}q^{\frac{r^{2}}{2}}e^{2\pi irz},\\
    \vartheta_{4}(\tau,z)&=\sum_{r\in\mZ}(-1)^{r}q^{\frac{r^{2}}{2}}e^{2\pi irz}.
\end{align}
Define $\vartheta_{j}(\tau,0)\equiv \vartheta_{j}(\tau)$ for $j=2,3,4$. Their transformations under the generators of $\rSL(2,\mathbb{Z})$ are
\begin{align}
    \vartheta_{2}(\tau+1)&=e^{2\pi i/8}\vartheta_{2}(\tau),\qquad\vartheta_{2}\left(-1/\tau\right)=\sqrt{-i\tau}\,\vartheta_{4}(\tau),\\
    \vartheta_{3}(\tau+1)&=\vartheta_{4}(\tau),\qquad \qquad \,\, \vartheta_{3}\left(-1/\tau\right)=\sqrt{-i\tau}\,\vartheta_{3}(\tau), \label{eq:theta3trafos}\\
    \vartheta_{4}(\tau+1)&=\vartheta_{3}(\tau),\qquad \qquad \,\, \vartheta_{4}\left(-1/\tau\right)=\sqrt{-i\tau}\,\vartheta_{2}(\tau).
\end{align}

Two useful identities for us are
\be 
\label{eq:thetaeta12eta24}
\begin{split} 
&\vartheta_3(2\tau)^2=2\,\frac{\eta(\tau+1/2)^{3}}{\vartheta_2(\tau+1/2)},\\
&\vartheta_3(2\tau)\,\eta(2\tau)=e^{-\pi i/12}\,\eta(\tau+1/2)^{2},
\end{split}
\ee 
which follow from the product representation for the Jacobi theta series. 
\subsection{Hardy-Ramanujan-Rademacher Formula for Fourier Coefficients}
\label{app:kloosterman}

We recall the Hardy-Ramanujan-Rademacher formula for the Fourier coefficients of vector-valued modular forms~\cite{Hardy:1918, Rademacher:1938, Dijkgraaf:2000fq}. Consider a weakly holomorphic vector-valued modular form $f_\mu(\tau)$, $\mu=1,\dots,D$ of negative weight $w<0$. The Fourier expansion of $f_\mu$ reads
\be
f_\mu(\tau)=\sum_{m\in \mathbb{N}} F_\mu(m-\Delta_\mu)\,q^{m-\Delta_\mu},\qquad q=e^{2\pi i\tau},
\ee 
with Fourier coefficients $F(m-\Delta_\mu)$ defined through
\be 
\label{eq:FC}
F_\mu(m-\Delta_\mu)=\int_{iY}^{iY+1} f(\tau)\,q^{-m+\Delta_\mu}\,d\tau,\qquad Y\in \mathbb{R}_+.
\ee 
The function $f_\mu$ transforms under modular transformations
\be
\label{eq:fmutrafo}
f_\mu\left(\gamma(\tau)\right)=(c\tau+d)^w \sum_{\nu=1}^D M(\gamma)_\mu^\nu\,f_\nu(\tau),
\ee 
with 
\be 
\label{Defgamma}
\gamma=\left(\begin{array}{cc} a & b \\ c & d  \end{array}\right)\in \rSL(2,\mathbb{Z}).
\ee
We have the following useful relations for $M$,
\be 
\label{eq:Mids}
\begin{split}
&M^{-1}(\gamma)=M(\gamma^{-1}),\\
&M(-\gamma)=M(\gamma)\,e^{\pi iw},\qquad c>0.
\end{split}
\ee 

For $m-\Delta_\mu\geq 0$, an exact formula for the Fourier coefficients can be derived using the an integration contour over Ford circles. These are an infinite set of circles anchored at the Farey fractions in the interval $(0,1]$. One could equivalently work with the Farey fractions in the interval $[0,1)$. The fractions are labeled by $-d/c$ with $c$ and $d$ relatively prime integers. The Fourier coefficients are then given in terms of the Kloosterman sum $K_c$ and Bessel function $I_\nu$ by,
\be 
\label{FourierRadem}
\begin{split}
    F_{\mu}(m-\Delta_\mu)&=2\pi\sum_{n-\Delta_{\nu}<0}F_{\nu}(n-\Delta_\nu)\sum_{c=1}^\infty\frac{1}{c}K_{c}(m-\Delta_{\mu},n-\Delta_{\nu}) \\
    &\times\left(\frac{|n-\Delta_{\nu}|}{m-\Delta_{\mu}}\right)^{\frac{1-w}{2}}I_{1-w}\left(\frac{4\pi}{c}\sqrt{(m-\Delta_{\mu})|n-\Delta_{\nu}|}\right).
\end{split}
\ee 
We have for $\Gamma(s)$,
\be 
\Gamma(s)=\left\{\begin{array}{rr} \frac{(2n)!}{4^n\,n!} \sqrt{\pi}, & \qquad s=1/2+n, n\in \mathbb{N},\\ 
(s-1)!, & \qquad s\in \mathbb{N}^{*}.\end{array}\right.\label{Gammafunction}
\ee 

The Bessel function $I_{\nu}(z)$ and Kloosterman sum $K_{c}(m-\Delta_{\mu},n-\Delta_{\nu})$ are defined as follows:
\begin{align}
    &I_{\nu}(z)=\left(\frac{z}{2}\right)^{\nu}\sum_{k=0}^{\infty}\frac{(\frac{1}{4}z^{2})^{k}}{k!\,\Gamma(\nu+k+1)},\\
    &K_{c}(\delta_{\mu},\delta_{\nu})=i^{-w}\sum_{-c\leq d< 0\atop (c,d)=1}{{M^{-1}(\gamma)}^{\nu}}_{\mu}
    \exp{\left[2\pi i\left(\delta_{\nu}\frac{a}{c}+\delta_{\mu}\frac{d}{c}\right)\right]}.\label{KloosterSum}
\end{align}
The behavior of the Bessel function $I_\nu$ for large argument implies that for large $m$, 
\be
\label{eq:logF}
\log(F_\mu(m))\simeq 4\pi \sqrt{m\Delta_{\nu,{\rm max}}},
\ee 
with $\Delta_{\nu,{\rm max}}$ the maximal value among the $\Delta_\nu$. The magnitude of the Kloosterman sum is bounded by $c^{1-\epsilon}$ for a sufficiently small $\epsilon$. More stringent bounds are obtained by Kloosterman and Weil. 

Eq. \eqref{FourierRadem} has a smooth limit for $m-\Delta_{\mu}\to 0$, since the vanishing denominator on the second line is cancelled by the Bessel function in the limit, $\lim_{z\rightarrow 0}I_{\nu}(z)\rightarrow\left(\frac{z}{2}\right)^{\nu}\frac{1}{\Gamma(\nu+1)}$. Therefore for $\Delta_{\mu}\in \mathbb{N}$, the constant term of $f_\mu$ is given by,
\begin{align}
\label{eq:FConst}
    F_{\mu}(0_\mu)=2\pi \sum_{n-\Delta_{\nu}<0}\frac{(2\pi|n-\Delta_{\nu}|)^{1-w}}{\Gamma(2-w)}F_{\nu}(n-\Delta_\nu)\sum_{c=1}^{\infty}c^{w-2}K_{c}(0_\mu,n-\Delta_{\nu}).
\end{align}
We note that the requirement for the weight $w<0$ is important here. While the Rademacher method has been developed for the coefficients of weakly holomorphic modular forms with weight $w=0$, the Circle Method does in general not converge to the constant term $F_{\mu}(0_\mu)$ of the modular form due to non-vanishing error terms \cite{Rademacher:1938}.

\section{Generalized Exponential Integrals}
\label{App:C}
The exponential integral ${\rm Ei}(x): \mathbb{R}\backslash \{0\}\to \mathbb{R}$ is defined as~\cite{Whittaker1996, Abramowitz1965}
\be 
\label{DefEi}
\text{Ei}(x)=-\int_{-x}^\infty e^{-t}t^{-1}dt,\quad\forall\  x>0.
\ee 
For $x>0$, ${\rm Ei}(x)$ should be understood as a principal value due to the singularity at $t=0$. 
For $x>0$, it can be written in the form
\be 
\label{Eix}
{\rm Ei}(x)=2\int_0^x \frac{\sinh(t)}{t}dt-\int_x^\infty e^{-t} t^{-1} dt.
\ee 
The integrand of the first integral has a smooth limit $t\to 0$.

For $z\in \mathbb{C}^*$, we furthermore recall the definition of the generalized exponential integral $E_s(z):\mathbb{C}^*\to \mathbb{C}$, also given in the main text as Eq. (\ref{DefExpI})~\cite[Eq. (8.19.2)]{DLMF},
\begin{align}
\label{DefExpIApp}
    E_{s}(z)=\left\{\begin{array}{cc}
        z^{s-1}\int_{z}^{\infty}e^{-t}t^{-s}dt, & \textrm{for }z\in\mC^{*}, \\
        \frac{1}{s-1}, & \textrm{for }z=0,\ s\neq 1, \\
        0,&\textrm{for }z=0,\ s=1.
    \end{array}\right.%\label{generalizedexp}
\end{align}
The integral for $s>1$ is defined through analytic continuation as discussed below. 

For $x\in \mathbb{R}_+$, we have $E_1(x)=-{\rm Ei}(-x)$. Integral shifts of $s$ are related through partial integration,
\be 
\label{ExpIPI}
e^{-z}=z\,E_s(z)+s\,E_{s+1}(z),\qquad \left(\frac{d}{dz}\right)^{(s-1)}E_{s}(z)=(-1)^{s-1}E_{1}(z).
\ee 
The recursion formula can be solved for integer $s\geq 1$,
\be
E_s(z)=e^{-z}\sum_{\ell=0}^{s-2} \frac{(s-\ell-2)!}{(s-1)!}\,(-z)^\ell+\frac{(-z)^{s-1}}{(s-1)!}\,E_1(z).
\ee 
For $z\in \mathbb{R}^+$, $E_{s}(z)>0$, such that Eq. (\ref{ExpIPI}) gives an upperbound for $E_s$,
\be
E_s(z) < \frac{e^{-z}}{z},\qquad z \in \mathbb{R}^+, \,\, s>0. \label{equa:geiupperbound}
\ee 

For $s<1$, $E_s(z)$ is regular around $z=0$. For $s\geq 1$, $z=0$ is a branch point. We choose the branch cut along the negative real axis, $z\in \mathbb{R}^-$. The discontinuity across the branch cut is
\be
\lim_{\delta \downarrow 0} \left(E_{s}(-x+i\delta)-E_{s}(-x-i\delta)\right)=-\frac{2\pi i\, x^{s-1}}{\Gamma(s)}=x^{s-1}\int_{\mathcal{H}}e^{-t}(-t)^{-s}dt,\label{equa:Estrespass}
\ee
where $x\in \mathbb{R}_+$, and the final expression is the Hankel representation of the Gamma function.

To avoid ambiguity, we will assume in the following that the contour of the exponential integral $E_s(-x)$ is infinitesimally deformed into the lower half plane. With $E_s(z)=E_s(z^*)^*$, we then find that the discontinuity gives the imaginary part of $E_s$,\footnote{NB Some other places in the literature and also Mathematica define $E_s(-x)$ as the limit from the upper-half-plane, $\delta\uparrow 0$, which gives the opposite sign for ${\rm Im}(E_{s}(-x))$.}
\be
\label{eq:ImEs}
{\rm Im}(E_{s}(-x))=\lim_{\delta\downarrow 0}
{\rm Im}(E_s(-x-i\delta))=\frac{\pi \, x^{s-1}}{\Gamma(s)},\qquad s\geq 1.
\ee 
For $E_1(-x)$ with $x\in \mathbb{R}_+$, we then have
\begin{align}
    E_{1}(-x)=\lim_{\delta\downarrow 0}E_{1}(-x-i\delta)=-\textrm{Ei}(x)+\pi i.
\end{align}
and for $E_2$,
\be
E_2(-x)=e^x-x\, {\rm Ei}(x)+\pi i x.
\ee 

For $z\neq 0$, $E_s(z)$ is related to the incomplete Gamma function,
\be 
\Gamma(\alpha,z)=\int_z^\infty e^{-t}t^{\alpha-1}\, dt,
\ee 
as
\be 
E_s(z)=z^{s-1} \Gamma(1-s,z).
\ee

 \end{document}